\documentclass[11pt]{aastex}
\usepackage{emulateapj5,apjfonts}
\usepackage{onecolfloat}
\usepackage{epsf}

\shortauthors{Bell et al.\ 2001}
\shorttitle{The Effects of Dust}

\newcommand{\ha}{{\rm H$\alpha$ }}
\newcommand{\hb}{{\rm H$\beta$ }}
\newcommand{\hbns}{{\rm H$\beta$}}
\newcommand{\hahb}{{\rm H$\alpha$/H$\beta$ }}
\newcommand{\hahbns}{{\rm H$\alpha$/H$\beta$}}

\newcommand{\hans}{{\rm H$\alpha$}}
\newcommand{\hii}{{\rm H\,{\sc ii} }}
\newcommand{\fir}{{\rm FIR }}
\newcommand{\firns}{{\rm FIR}}
\newcommand{\fuv}{{\rm FUV }}
\newcommand{\fuvns}{{\rm FUV}}

\newcommand{\peg}{{\sc P\'egase }}
\newcommand{\arad}{{$A_{\rm H\alpha,Radio}$ }}
\newcommand{\abalm}{{$A_{\rm H\alpha,Balmer}$ }}
\newcommand{\aradns}{{$A_{\rm H\alpha,Radio}$}}
\newcommand{\abalmns}{{$A_{\rm H\alpha,Balmer}$}}
\newcommand{\afif}{{$A_{\rm 1500,FR}$ }}
\newcommand{\afifns}{{$A_{\rm 1500,FR}$}}

\slugcomment{{\sc To appear in ApJ: } January 20 2002, vol. 565, no. 1}

\begin{document}


\def\head{

\title{The effects of dust in simple environments: Large Magellanic
	Cloud \hii regions} 

\author{Eric F. Bell, Karl D.\ Gordon, Robert C. Kennicutt, Jr., \& 
	Dennis Zaritsky}
\affil{Steward Observatory, University of Arizona,
   Tucson, AZ 85721}
\email{(ebell,kgordon,robk,dennis)@as.arizona.edu}

\begin{abstract}
We investigate the effects of dust on Large Magellanic Cloud (LMC) \hii region
spectral energy distributions using 
arcminute-resolution far-ultraviolet (\fuvns), \hans, 
far-infrared (\firns), and radio images.
Widely-used indicators of the amount of light lost 
to dust (attenuation) at \ha and 
in the \fuv correlate with each other,
although often with substantial scatter. There are two interesting
systematic discrepancies.  First,
\ha attenuations estimated from the Balmer decrement are lower than those
estimated from the \hans-to-thermal radio luminosity ratio.
Our data, at this stage, cannot unambiguously 
identify the source of this discrepancy.
Second, the attenuation at 1500{\AA} and UV spectral 
slope, $\beta$, correlate, although the slope and scatter 
are substantially different from the correlation first derived for
starbursting galaxies by Calzetti et al. 
Combining our result with those 
of Meurer et al.\ for ultra-luminous
infrared galaxies and Calzetti et al.\ for starbursting galaxies, 
we conclude that no single relation between
$\beta$ and 1500{\AA} attenuation is applicable to all
star-forming systems.

\end{abstract}

\keywords{dust, extinction --- \hii regions --- 
Magellanic Clouds --- galaxies: individual (LMC) --- ultraviolet: galaxies}
}

\twocolumn[\head]

\section{Introduction} \label{sec:intro}

Estimating where and how quickly stars are forming in galaxies
of different types, as a function of redshift, gives us direct
insight into where, when and how galaxies evolve 
\citep[see e.g.][for attempts to address this problem in a global sense]{madau96,blain99,steidel99}.  In this context,
understanding the systematic uncertainties plaguing SFR estimators
is of special importance.  One of the principal 
uncertainties of this type is the correction for emergent flux lost due
to dust. In a systematic effort to address this problem,
we begin by studying the effects of dust in the simplest of all star formation 
environments, \hii regions.

\begin{figure}[tbh]
\epsfxsize=\linewidth
\epsfbox{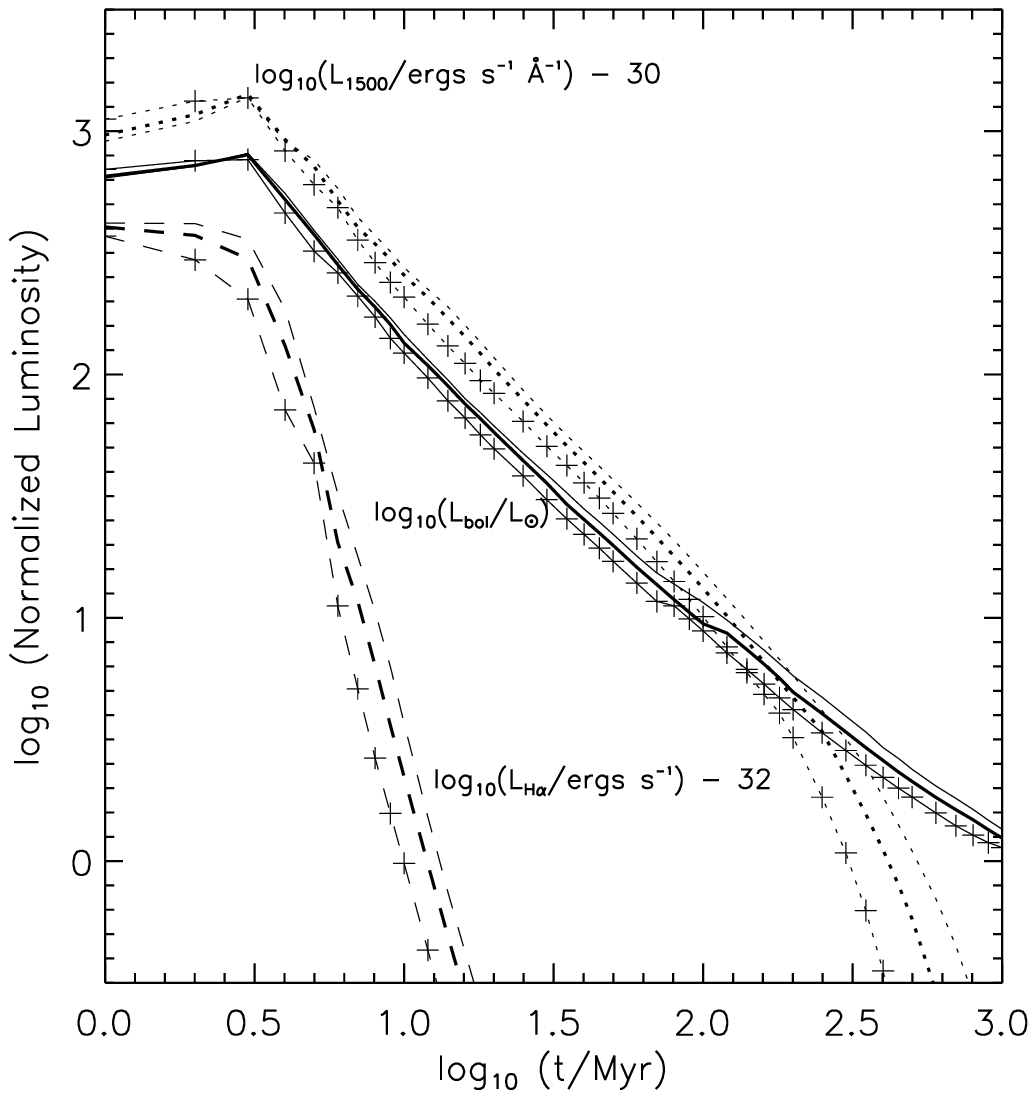}
\caption{\label{fig:indic} Age and metallicity dependence of 
the bolometric luminosity (L$_{\rm bol}$; {\it solid lines}), 
1500{\AA} luminosity (L$_{1500}$; {\it dotted lines}), and
\ha luminosity (L$_{\rm H\alpha}$; derived assuming
0.45 \ha photons per Lyman continuum photon; {\it dashed lines}) of a
1M$_{\sun}$ stellar population, as derived using the 
\peg stellar population synthesis
models \protect\citep{fioc01}.  Thick
lines denote the Z=0.008 (40\% solar) models, thin lines
denote the Z=0.004 (20\% solar) models, and thin lines
with crosses denote the Z=0.02 (solar) models.  All luminosities
at ages $\ga 3$ Myr decrease 
with increasing metallicity, primarily due to increased
stellar opacities.  At younger ages, the behavior is more complex
because of the interplay of mean molecular weight and opacity effects.
}
\end{figure}

Before discussing the effects of dust, it is necessary to 
understand the nature and origins of the different SFR indicators.
SFRs are usually estimated from spectral regions
where the energy output from young stars dominates.
Massive, short-lived stars produce copious quantities of 
ultraviolet (UV) radiation.  This UV light may be directly observed, or 
dust may absorb much of it,
reprocessing it into the far-infrared (\firns).
In the limit of dust absorbing all of the energy output of 
young stellar populations, the \fir radiation is approximated
by the bolometric luminosity.  The time evolution of the 
UV and bolometric luminosities
is shown in Fig.\ \ref{fig:indic} for a single burst from the \peg stellar
population models \citep{fioc01}: 
the UV and bolometric luminosities evolve similarly
for more than 100 Myr.  The most massive and short-lived
stars produce radiation capable of ionizing the hydrogen gas
in their natal clouds, producing a \hii region.  
This ionized hydrogen plasma produces
line emission (the Balmer series is the most frequently observed
series of emission lines) and thermal bremsstrahlung in the radio:
these radiations evolve much more quickly than the 1500{\AA} UV or 
bolometric luminosities (Fig.\ \ref{fig:indic}).

However, the \ha and UV luminosities will be
strongly reduced by dust.  
The amount of light lost from a system is referred to 
as the {\it attenuation}. The attenuation and the more commonly used
{\it extinction} are the same for a single star with dust along
the line-of-sight.  However, for more complex
systems, the attenuation is a function of the geometry of both
the dust and stars.  Classic methods that work for determining
extinctions (e.g. color excesses) will not reliably determine
attenuations.

Despite the difficulties introduced by geometry, 
it is possible to estimate the amount of 
luminosity lost at \ha or in the UV, allowing the true SFR to be
estimated.  Some attenuation estimators rely solely on UV/optical data. 
For example, it is common to estimate the amount
of \ha light lost to dust by comparing the \ha and \hb emission-line fluxes,
which have an intrinsic ratio fixed by atomic physics in most astrophysical
environments.  The observed relative ratio therefore gives an 
estimate of the reddening: this can be transformed into an 
attenuation under the assumption of a particular
attenuation curve \citep[see e.g.][for a test of this method, 
and references therein]{quillen01}.
A common approach to estimate the attenuation of UV stellar light
is to use the UV spectral slope $\beta$.  This approach,
pioneered by \citet{calzetti94} and \citet{meurer99}, uses
the relatively narrow range in intrinsic $\beta$ for young, 
luminous stellar populations \citep[$-2.6 < \beta < -1.5$ including
even metal-rich single bursts with ages younger than 20 Myr;][]{sb99}
to estimate the reddening. This reddening is observed
to correlate with UV attenuation with only modest scatter for 
local starbursting
galaxies \citep{calzetti94,heckman98,meurer99}. However, if one applies
this relationship to systems other than those for which it was
calibrated, then one is assuming that the relation holds over a variety
of star-forming systems. 

Other attenuation estimators rely on data from 
beyond UV and optical wavelengths. For example, the \ha attenuation
can be estimated from the ratio of \ha to thermal radio continuum 
\citep[e.g.][]{israel80,k83,vdh88}. In this case, one concern is that
the thermal radio emission may be 
contaminated by non-thermal emission. This contamination is severe
for most galaxies \citep[in fact, the non-thermal radio may 
correlate with SFR, although we do not discuss it in this 
paper; see e.g.][]{condon92,cram98}.  Another such attenuation indicator, 
this time for the far-UV (\fuvns), uses the ratio of \fuv to \firns, which 
measures the amount of \fuv light reprocessed into the 
\fir \citep[e.g.][]{buat96,meurer99,fluxrat}. 

There are inevitable 
limitations to any attenuation estimator.  Bearing in
mind the importance of constructing 
relatively accurate, unbiased SFRs as a probe of galactic evolution,
it is vital to critically assess the applicability 
and efficacy of attenuation estimators. Considering
data from as many wavelengths as possible is an 
important part of this assessment: 
by using UV, \hans, \firns, and thermal radio luminosities it is 
possible to construct different estimates of the attenuation
at a given wavelength.  Examples of this kind of multi-wavelength
approach for galaxies include \citet{calzetti94}, \citet{meurer99}, 
\citet{sullivan00}, and \citet{uit}.

An essential test of the robustness of attenuation indicators is their
behavior in the simplest of all composite star-forming systems:
\hii regions.  Without a firm understanding of the attenuation 
of \ha and \fuv light 
in \hii regions, we cannot claim a firm understanding
of the effects of dust on galactic scales.  
Because \hii regions have ionized gas and 
relatively well-constrained ages (the dominant
populations have to be capable of ionizing hydrogen), it is possible to  
intercompare the attenuations derived for the ionized gas and 
those derived for the stellar population inside the \hii region.
We have chosen to 
use the sample of Large Magellanic Cloud (LMC) \hii regions
from \citet{cap86} for this
purpose.  This sample of \hii regions has a number of advantages
for this purpose: a common distance,
modest galactic foreground 
extinction, and the availability of large-format images and published
data in the UV, optical, \firns, and radio spectral regions with a
useful physical resolution (we adopt 
4.9$\arcmin$ [80 pc] diameter apertures for this work).

The remainder of this paper is organized as follows.
\S\ref{sec:tut} discusses the construction and limitations
of \fuv and \ha attenuation indicators.
\S\ref{sec:data} describes the multi-wavelength imaging data, the 
morphology of the LMC at different wavelengths, and the 
photometry of individual \hii regions.  \S\ref{sec:dust}
quantifies the relationships between 
different \hii region attenuation indicators. 
In \S\ref{sec:disc}, we discuss in detail two particular comparisons: 
the well-documented discrepancy between \ha attenuations estimated
from Balmer decrements and \hans-to-radio ratios, and the 
highly scattered and rather steep correlation between 
\fuv attenuation and UV spectral slope $\beta$.
\S\ref{sec:conc} summarizes our conclusions from this study.
We assume a distance
to the LMC of 45.9$\,\pm\,$2 kpc \citep[e.g.][]{fitz01},
which only affects the absolute scaling of the SEDs presented in 
\S\ref{sec:data} because the attenuations and flux ratios are 
distance-independent.  In correcting our attenuation
estimates for foreground galactic extinction, we adopt a foreground
reddening of $E(B-V) = 0.06$, following \citet{sfd} and \citet{oestreicher95}.

\section{Quantitative Attenuation Indicators} \label{sec:tut}

\begin{figure*}[tbh]
\hspace{3.75cm}
\epsfxsize=10.5cm
\epsfbox{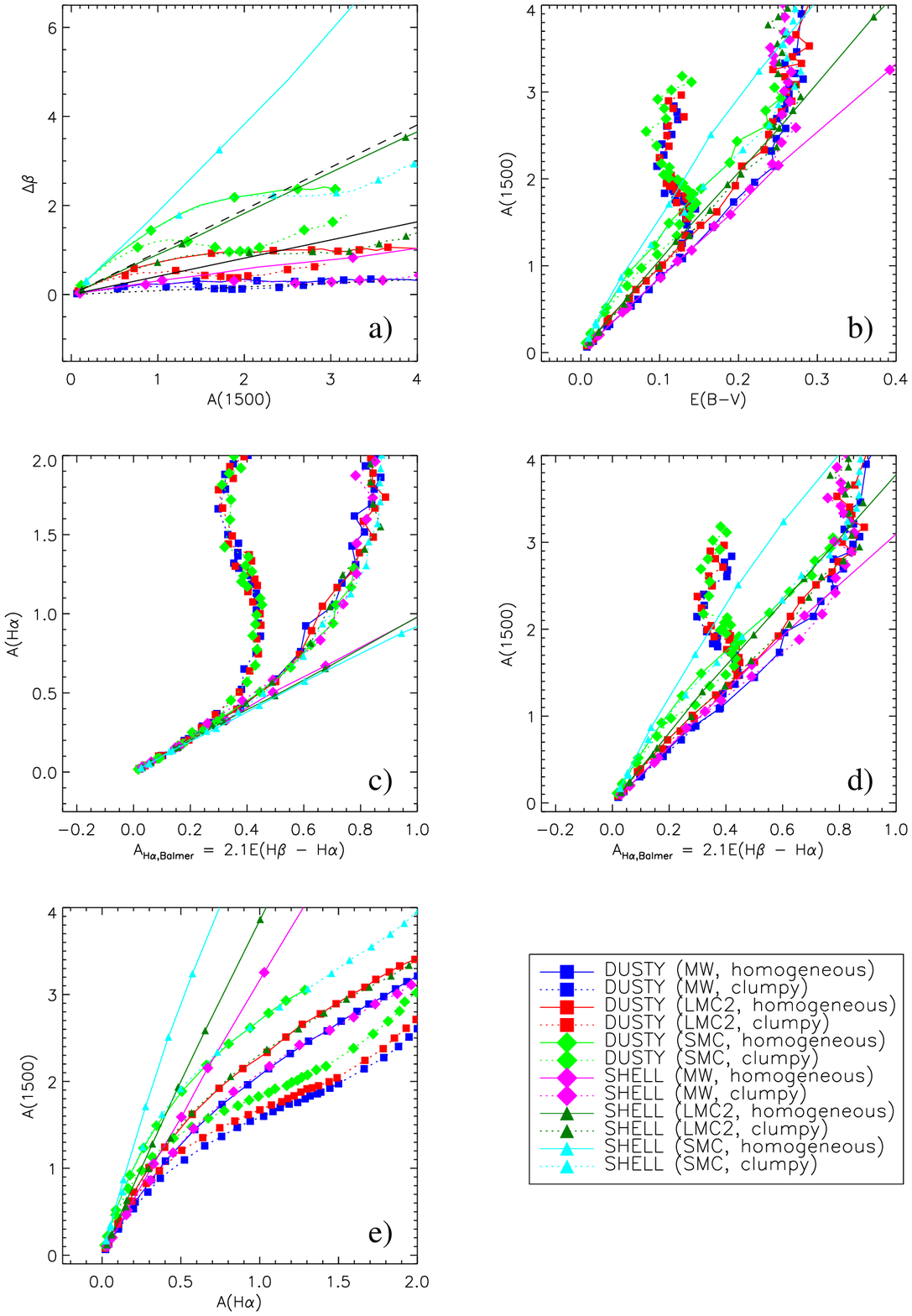}
\caption{\label{fig:radtrans}
Predictions from radiative transfer models regarding the 
relationships between different attenuation indicators.
{\it (a)} 1500{\AA} attenuation against change in UV spectral slope $\Delta\beta$,
{\it (b)} $E(B-V)$ against 1500{\AA} attenuation,
{\it (c)} Balmer-derived \ha attenuation against true \ha attenuation, 
{\it (d)} Balmer-derived \ha attenuation against 1500{\AA} attenuation, and
{\it (e)} true \ha attenuation against 1500{\AA} attenuation.
Models are described in the legend, and expanded on in 
the text.  Panel {\it (a)} also shows screen models with LMC (solid line)
and SMC Bar (dashed line) extinction curves.
}
\end{figure*}

In this section, we discuss the estimation of \ha and UV attenuations
from multiwavelength data.  This serves a dual purpose: to introduce 
the methods used to calculate
the attenuation indicators later in the paper, and to instill
an intuition for what to expect when
comparing the different attenuation indicators, including 
the nontrivial effects of radiative transfer.

\subsection{Attenuation of far-UV light}

Using measured UV reddenings to estimate the \fuv attenuation is becoming
increasingly fashionable due to its observational efficiency.
The UV continuum spectral slope $\beta$, defined by:
\begin{equation}
F_{\lambda} \propto \lambda^{\beta},   \label{eqn:beta}
\end{equation}
where $F_{\lambda}$ is the flux per unit wavelength $\lambda$, is
used to estimate
the attenuation in the UV \citep[e.g.][]{calzetti94,meurer99,adel00}
because one presumes to understand the color of the underlying starlight,
and so any deviation from this color is attributed to dust. 
To the degree that one believes the underlying assumptions in 
this method, one could, for example, 
also use optical reddenings to estimate UV attenuation,
or use UV reddenings to estimate the optical attenuation. Clearly, such
an exaggerated extension of the technique loses credibility and one begins to 
wonder where, exactly, the technique may break down.
The relationship between the UV or optical reddenings
and the \fuv attenuation is non-trivial, as illustrated
in panels {\it (a)} and {\it (b)} of Fig.\ \ref{fig:radtrans}.

In Fig.\ \ref{fig:radtrans}, we show predictions from 
the radiative transfer model {\sc dirty} \citep{clump2,dirty1,dirty2}.
The relationships between different attenuation indicators are 
shown under different assumptions about the extinction curve
(MW$=$Milky Way, LMC2$=$a supershell near 30 Dor, 
SMC$=$Small Magellanic Cloud Bar),
dust clumping (homogeneous or clumpy, with a density ratio of
100), and geometry (dusty$=$uniformly mixed stars and dust, 
shell$=$inner dust-free region of stars surrounded by a shell
of star-free dust).  
The main point to note from the
panels {\it (a)} and {\it (b)} of Fig.\ \ref{fig:radtrans}
is that it is possible to have a large range in 
\fuv attenuation for a given UV or optical reddening, depending
on the dust/star geometry and/or shape of extinction curve.
Bearing in mind the extensive empirical evidence for a highly
environmentally-sensitive UV extinction curve 
\citep[e.g.][]{fitz86,gordon98,misselt99,clayton00},
and the diversity of dust/star geometries observed in many astrophysical
systems, our expectation is that 
the validity of UV or optical reddening-based attenuations
is open to suspicion in any but the best-studied and characterized cases.

It is possible to avoid the need to assume a particular
extinction curve in deriving the \fuv attenuation by comparing 
the observed \fuv starlight with the \fir luminosity.  Because the 
majority of the bolometric flux of a young stellar population 
is emitted in the \fuv and because the \fir flux is dominated
by the re-radiation of attenuated UV light, this comparison
provides an estimate of the \fuv 
attenuation that is insensitive to the extinction curve
and geometry when averaged over $4\pi$ steradians
\citep[see Figs.\ 1 and 2 of][]{fluxrat}.  We use an IDL
subroutine, available from K.\ D.\ G., and described fully
in \citet{fluxrat}, to estimate the attenuation at 1500{\AA}
and 1900{\AA} from the 1500{\AA}, 1900{\AA}, and \fir fluxes.  This subroutine
takes into account the full uncertainties implied by differences
in dust geometry, dust type and stellar populations.  For 1500{\AA}
attenuations between 0 and 4 mags, 
the 1500{\AA} flux ratio method-derived attenuation can
be approximated by a modification of the simple energy-balance estimate
to within 0.03 mag:
\begin{equation}
A_{\rm 1500,FR} \sim 0.86 \left [ -2.5 \log_{10} \frac{\nu f_{\nu}}{\nu f_{\nu} + L_{\rm FIR}} \right ] - 0.1, \label{eqn:fr}
\end{equation}
where the bracketed quantity is the simple energy-balance estimate of 1500{\AA}
attenuation ($\nu f_{\nu}$ is evaluated at 1500{\AA}), 
$L_{\rm FIR}$ is an estimate of the total \fir luminosity
(estimated as described in \S \ref{sec:estimate}), 
and the scaling and subtraction approximately correct
for SED and attenuation curve shape.
This method has important limitations.  First,
it requires isotropy on scales larger than the 
illuminating source. An example of a geometry for which
this approach fails is the case of a torus of dust
outside a cluster of \fuvns-bright stars, in which the torus
is oriented perpendicular to 
the line of sight. Such an region would have a large \fir
luminosity but {\it no} attenuation along the line of sight.
Second, the method requires the young population to dominate the heating.
The method, some of its uncertainties, and its limitations are presented in 
\citet{fluxrat}: variants of this method were used by e.g.\ 
\citet{buat96} or \citet{meurer99} 
to explore the attenuations of spiral and starburst galaxies respectively.

\subsection{Attenuation of \ha light}

Estimating the \ha luminosity lost to dust is relatively
straightforward.  We discuss two methods: both
estimate only the attenuation of the \ha 
radiation and are insensitive to the amount of Lyman continuum
radiation lost to dust before being down-converted into \hans.
The first is based on the comparison of the \ha line luminosity to the thermal 
radio luminosity.  This ratio, in the absence of dust,
is given by equations A5 and
A10 of \citet{cap86}:  
\begin{eqnarray}
f_{\rm H\alpha}/f_{\rm 8.55 \,GHz} & = & 8.659 \times 10^{-9} \, {\rm erg\,cm^{-2}\,s^{-1}\,Jy^{-1}} \nonumber \\
& & \times \, (t_e)^{-0.44} \,[8.666 + 1.5\ln(t_e)]^{-1},
\end{eqnarray}
where $f_{\rm H\alpha}$ is the \ha flux in erg\,cm$^{-2}$\,s$^{-1}$, 
$f_{\rm 8.55 \,GHz}$ is the 8.55 GHz flux in Jy, and $t_e$ is the electron 
temperature in units of 10$^4$\,K.
The ratio depends weakly on 
electron temperature: we will adopt an electron temperature of 
9500\,K $\pm$ 1000\,K \citep{pagel78}, corresponding in dust-free
\ha to radio ratio uncertainties of $\sim$8\%.  
In practice, the thermal radio luminosity is used to construct an expectation
for the \ha line luminosity. Then, the observed \ha line luminosity is 
compared to the radio expectation to derive an effective attenuation
\aradns.
Although this should be a robust attenuation indicator, 
accurate {\it thermal} radio luminosities are difficult to measure, even
for individual \hii regions.

The second approach is to estimate the \ha attenuation using 
the observed \hahb ratio for a given \hii region.  Assuming
a standard R$_V = 3.1$ Milky Way extinction curve and a foreground
screen dust distribution, 
\citet{cap86} derive the following expression for \ha attenuation
estimated from the Balmer decrement \abalm
given an observed \hahbns:
\begin{equation}
A_{\rm H\alpha,Balmer} = 5.25 \log_{10} \left(\frac{{\rm H\alpha/H\beta}}
 {2.859\,t_e^{-0.07}}\right),
\end{equation} 
or, equivalently, $A_{\rm H\alpha,Balmer} = 2.1\,E({\rm H\beta - H\alpha})$.
\hahb is observationally straightforward to measure;
however, different dust extinction curves or geometries change 
the conversion between reddening and attenuation from the standard
R$_V = 3.1$ foreground screen.  

The sensitivity of these estimators to geometry and dust properties
is illustrated in panel {\it (c)} of Fig.\ \ref{fig:radtrans},
where \abalm is shown against the true \ha attenuation A(\hans).
For homogeneous shell or screen geometries, 
\abalm and A(\hans) are essentially identical; however,
clumpy shell and dusty geometries show a `saturation' for \abalm $\sim$0.8
and $\sim$0.4 mag respectively, where only the least reddened stars 
are observed, resulting in high attenuations but modest reddenings.

\subsection{Comparing \ha and far-UV attenuations}

The relationship between \ha and \fuv attenuations is complex
because it depends on extinction curve shape, dust clumping 
properties, and geometry 
(panel {\it (d)} of Fig.\ \ref{fig:radtrans}).
In particular, for clumped dust or 
mixed star/dust geometries, the relationship between \abalm and 
1500{\AA} attenuation is non-linear.  Panel {\it (d)} is 
quite similar to panel {\it (b)}; both \abalm and 
$E(B-V)$ are essentially color excesses, and behave quite similarly.
Also, as shown in panel {\it (e)}, 
the relationship between true \ha and 1500{\AA} attenuation
is non-linear for clumped and/or dusty geometries (in quite
a different way than the Balmer-derived \ha attenuation). While
the \fuv attenuation `saturates' at around 2 magnitudes, the 
\ha attenuation continues to increase (because the \fuv becomes optically
thick, but the \ha remains optically thin). 
A further complication, which is more of a concern when comparing 
\ha and \fuv attenuation in 
galaxies, is that the \ha and \fuv light can come from 
spatially distinct locations.  In \hii regions, this is less
of a concern, and we will use 
the inter-comparison of \ha and \fuv attenuations 
to test the reproducibility of these attenuation estimators.

\section{Data analysis} \label{sec:data}

\subsection{Images} \label{subsec:data:images}

We use \fuvns, \hans, \firns, and 8.55 GHz 
radio images of the LMC to investigate the spectral energy 
distributions (SEDs) and attenuations 
for \hii regions drawn from 
the catalogs of \citet{cap85,cap86}.  Accordingly, we have 
compiled an extensive library of LMC images with 
roughly arcminute resolution from the literature.  
We describe these images in turn 
below, and in Table \ref{tab:data}.

\begin{deluxetable}{cllcc}
\tablewidth{240pt}
\tablecaption{LMC Data \label{tab:data}}
\tablehead{\colhead{Band} & \colhead{$\lambda$} & \colhead{$\Delta\lambda$} 
     & \colhead{FWHM} & \colhead{Ref} \\
    & \colhead{[$\micron$]} & 
     \colhead{[$\micron$]} & \colhead{[$\arcsec$]} &  }
\startdata
Rocket\_UV1 & 0.1495 & 0.020 & 100 & 1 \\ 
Rocket\_UV2 & 0.1934 & 0.022 & 100 & 1 \\
$R$          & 0.66   & 0.15   & 72    & 2 \\
\ha         & 0.6571 & 0.0013 & 72    & 2 \\
{\it MSX} B3      & 8.3    & 3.36 & 45 & 3 \\
IRAS 12$\micron$ & 12 & 6.5 & 38 & 4 \\  
IRAS 25$\micron$ & 25 & 11 & 42 & 4 \\
IRAS 60$\micron$ & 60 & 40 & 75 & 4  \\
IRAS 100$\micron$ & 100 & 37 & 135 & 4 \\
8550 MHz   & $3.5\times 10^{4}$ & \nodata & 210 & 5 \\ 
\enddata
\tablerefs{(1) \citet{smi87}, (2) \citet{ken95},
   (3) http://www.ipac.caltech.edu/ipac/msx/msx.html, 
   (4) \citet{bra98}, (5) \citet{fil98}}
\end{deluxetable}

\noindent $\bullet$
The UV images come from a rocket launched from Australia
in 1977 (for a full description see \citet{smi87}).  The
angular resolution is $\sim$100{\arcsec} over most of the useful part 
of the image.  We convert the raw
digital units to MJy/sr  
using calibrations supplied by R.\ Cornett (1999, private
communication). We rule out offsets in 
calibration larger than 12\% at the 2$\sigma$ level by comparing to
aperture fluxes of 10 star forming regions from 
{\it Ultraviolet Imaging Telescope} images with an effective
wavelength of 1614{\AA}.

\noindent $\bullet$
The H$\alpha$ and $R$ band images were 
taken with the Parking Lot Camera \citep{bot86}, 
but with a Texas Instruments 800$\times$800 pixel CCD \citep{ken95}.  
We calibrated the \ha image by comparing to aperture
\ha fluxes from \citet{cap85}. Our
calibration is consistent with the calibration of the H$\alpha$ image given by
\citet{ken95} to within 10\%.  The $R$ band image has not been 
photometrically calibrated.

\noindent $\bullet$
The {\it Midcourse Space Experiment} ({\it MSX}) 
B3 (8.3$\micron$) data were downloaded from the IPAC {\it MSX} site
\citep[for a description of the instrument and the data see][]{price01}.  
We use the mosaic image, with 45{\arcsec} instead of 18{\arcsec}
resolution, as our work is limited by the lower resolutions 
of the images in other passbands.

\noindent $\bullet$
The IRAS images at 12$\micron$, 25$\micron$, 60$\micron$ and
100$\micron$, processed with a maximum entropy
algorithm to improve resolution \citep{bra98}, were provided 
by M.\ Braun (2000, private communication). We confirm the
validity of structures seen in the resolution-enhanced images
by comparing the processed IRAS
12$\micron$ image with the {\it MSX} 8.3$\micron$ image.  The
morphologies of the structures in the images are very similar.  
At faint levels, the processed 
image displays the characteristic striping of IRAS
data.

\noindent $\bullet$
The radio images were provided by M.\ Filipovi\'c (2000, private
communication) and are described in detail in \citet{fil98}.  
These data are calibrated to better than 15\%, when tested against
`aperture' 5 GHz radio fluxes from \citet{cap86}: before comparison,
the 5 GHz radio fluxes were scaled down by 5\% to 
account for wavelength differences, assuming purely thermal radio emission.

All the images were aligned (translation and rotation
only) to the $R$ band image, for which J.\ Parker
(1999, private communication) provided an astrometric solution.
At least 5--10 
stars and \hii regions were used as the basis of the
alignment, and alignment was accurate to better than 20$\arcsec$ 
in all cases.  After the alignment, the images were transformed to a common NCP
projection with 30{\arcsec} pixels.  All of the images, 
except for the \ha and $R$ band image,
are calibrated in units of MJy/sr.
The \ha image is calibrated 
in ergs\,s$^{-1}$\,cm$^{-2}$ and, as stated before, the $R$ band image
is uncalibrated.  The original and calibrated, aligned images
are available from the authors.

\subsection{Morphology of the LMC at different wavelengths} 
  \label{subsec:data:morph}

\begin{figure*}[tbh]
\caption{\label{fig:image}
The morphology of the LMC at \hans, 1900{\AA}, $R$ band, and in the 
\fir 60$\micron$ band.  The 52 \hii regions in our sample are circled.}
\end{figure*}

Before discussing the SEDs and attenuations of the LMC \hii regions,
we must comment on one of the most striking aspects of 
Fig.\ \ref{fig:image} --- the strong difference between the 
distributions of the young stars (as probed by the \hans,
1900{\AA} or \fir emission) and older stars (as probed by the $R$ band).
The older stellar populations are strongly concentrated
in the bar, with an underlying disk-like envelope.  In contrast,
the younger stars have a clumpy distribution, and are
predominantly concentrated in the northern half of the LMC.
This interesting aspect of the data is beyond the scope of the
current work.

More subtle, but perhaps equally significant, differences are also 
present.  In \S \ref{sec:intro}, we saw that the UV and \fir
luminosities should track each other quite closely if
the amount and distribution of dust around young
stars does not depend on stellar age.  Therefore, a na\"{\i}ve
expectation might be that the morphologies of the LMC in the UV and \fir 
should be similar. In contrast, the \ha morphology could either
be similar or not depending on the detailed star formation history
over timescales of $\sim$10 to 100 Myr.
Surprisingly, the observation is that 
the \fir and \ha morphologies are similar (cf.\
Fig.\ \ref{fig:image}), and that they
differ from the morphology in the UV. 

This departure from the
na\"{\i}ve model has two interesting implications. 
Firstly, the assumption used in linking
UV to \fir luminosities has broken down. The similarity between
the \ha and \fir morphologies implies that there is a good
correspondence between the distribution of very young (ionizing)
stars and dust. However, as stars age (so that they are now no longer
strongly ionizing stars but still have relatively large UV fluxes), 
the distribution of dust must decouple from that of the stars.  
We suggest the simple 
scenario in which young, dusty \hii regions are dispersed by the 
cumulative effects of stellar winds and supernovae. This dispersal
(1) causes many of the \fuvns-bright regions to be faint in the \firns,
(2) explains the existence of large, extended
\fuvns-bright regions ringed by \ha and \fir emission from recent
star formation, that are triggered by the winds from the OB association in 
the center (e.g.\ Constellation {\sc iii}, in the top left quadrant
of the images), and (3) is consistent with other empirical evidence 
\citep[e.g.][and references therein]{charlot00}.

Secondly, the dissimilarity between the \ha and UV morphologies of the LMC
stresses a fundamental limitation of this work in its application
to the global situation in galaxies.
Because much of the UV flux of the LMC comes
from outside the \hii regions we study here, we cannot meaningfully constrain
the effects of dust on the total UV emission from the LMC (we will
address this issue in future work).  
The dichotomy between the sites of \ha and UV emission
also leads us to caution against attempts to link the effects of dust 
on galactic \ha and UV emission too closely: it may be 
impossible to meaningfully link the \ha and \fuv extinction properties of 
star-forming galaxies in any more than a statistical sense.

\subsection{SEDs} \label{subsec:data:seds}

\begin{figure*}[tb]
\hbox{%
\epsfxsize=8.8cm
\epsfbox{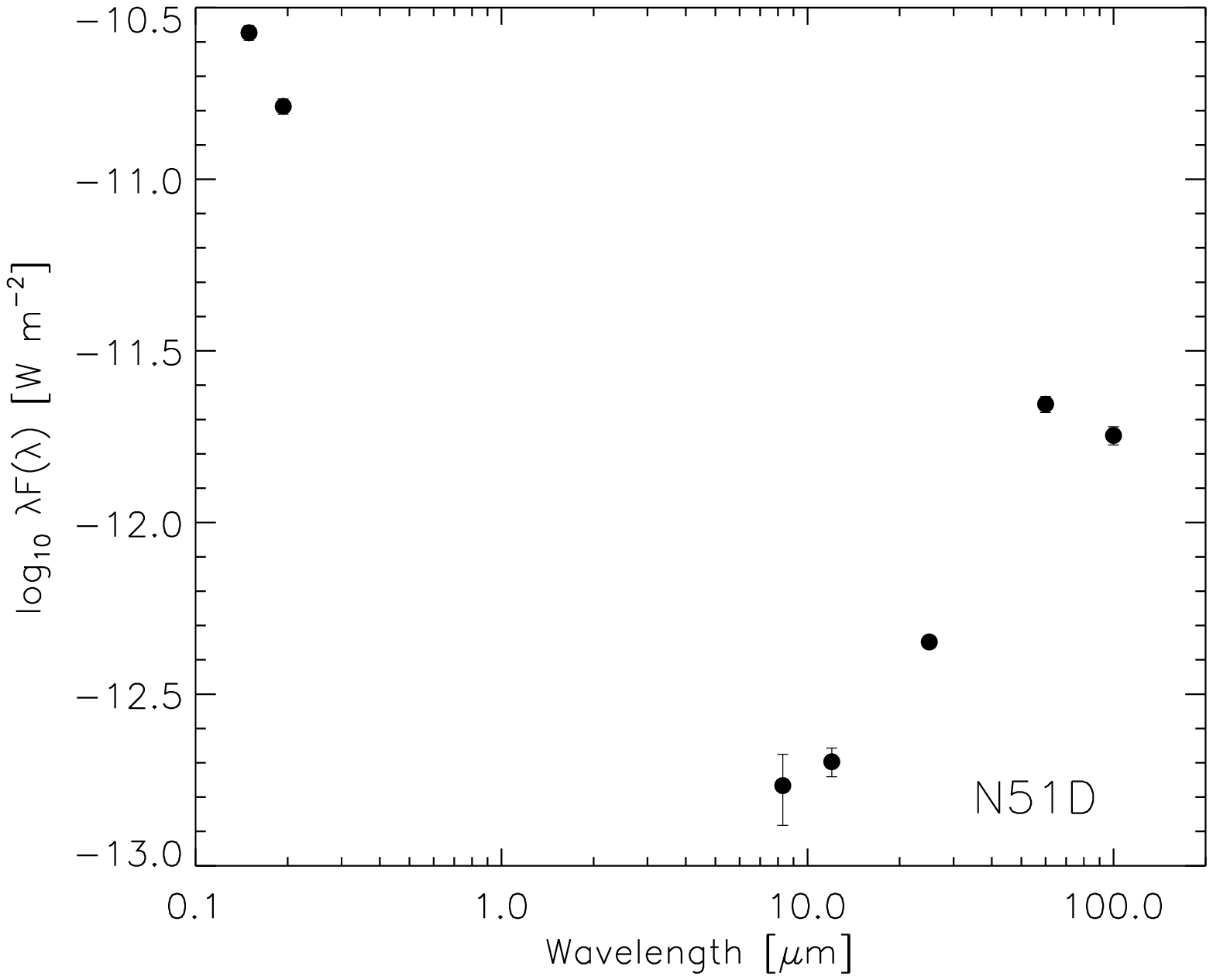}
\hfill
\epsfxsize=8.8cm
\epsfbox{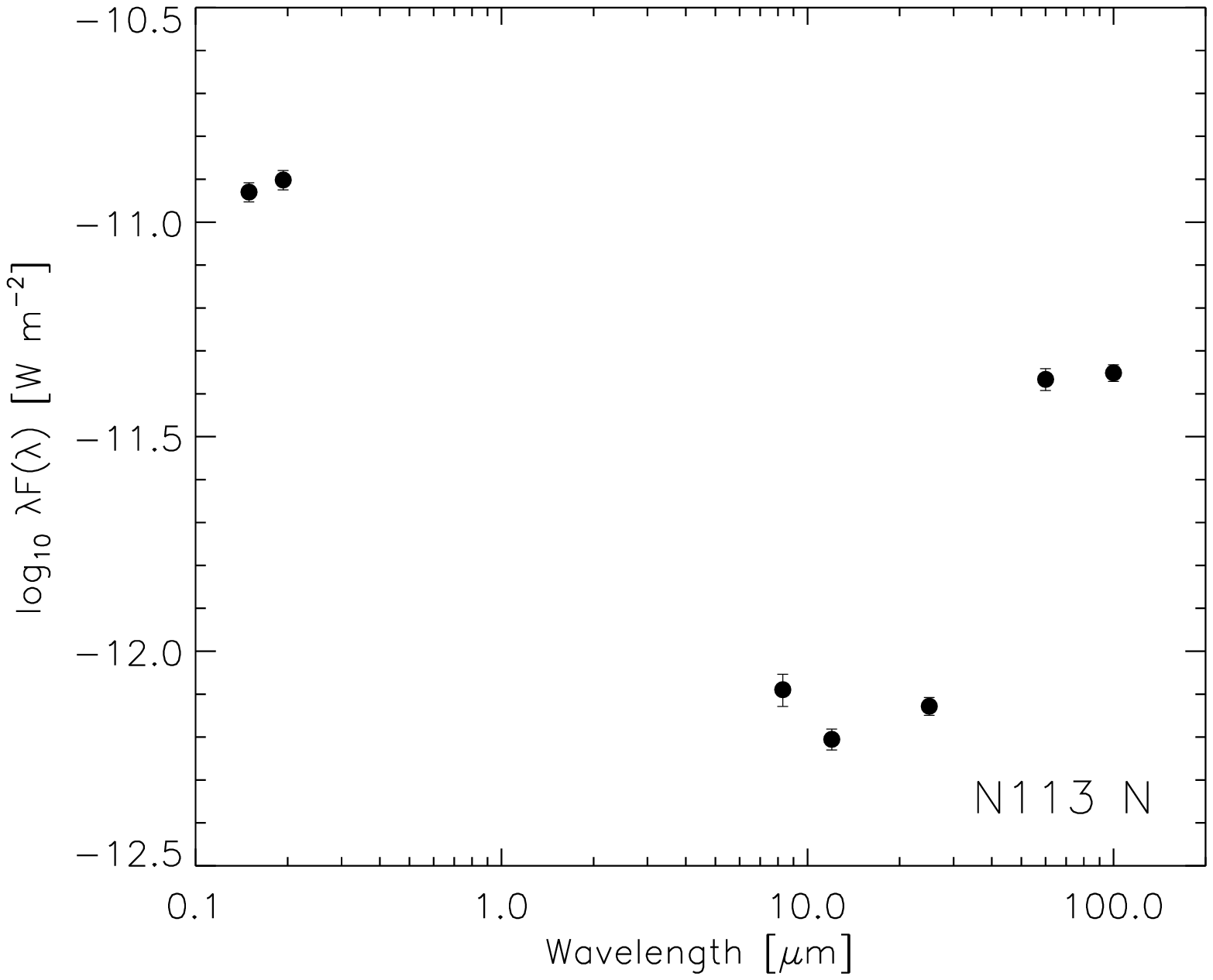}}
\vfill
\hbox{%
\epsfxsize=8.8cm
\epsfbox{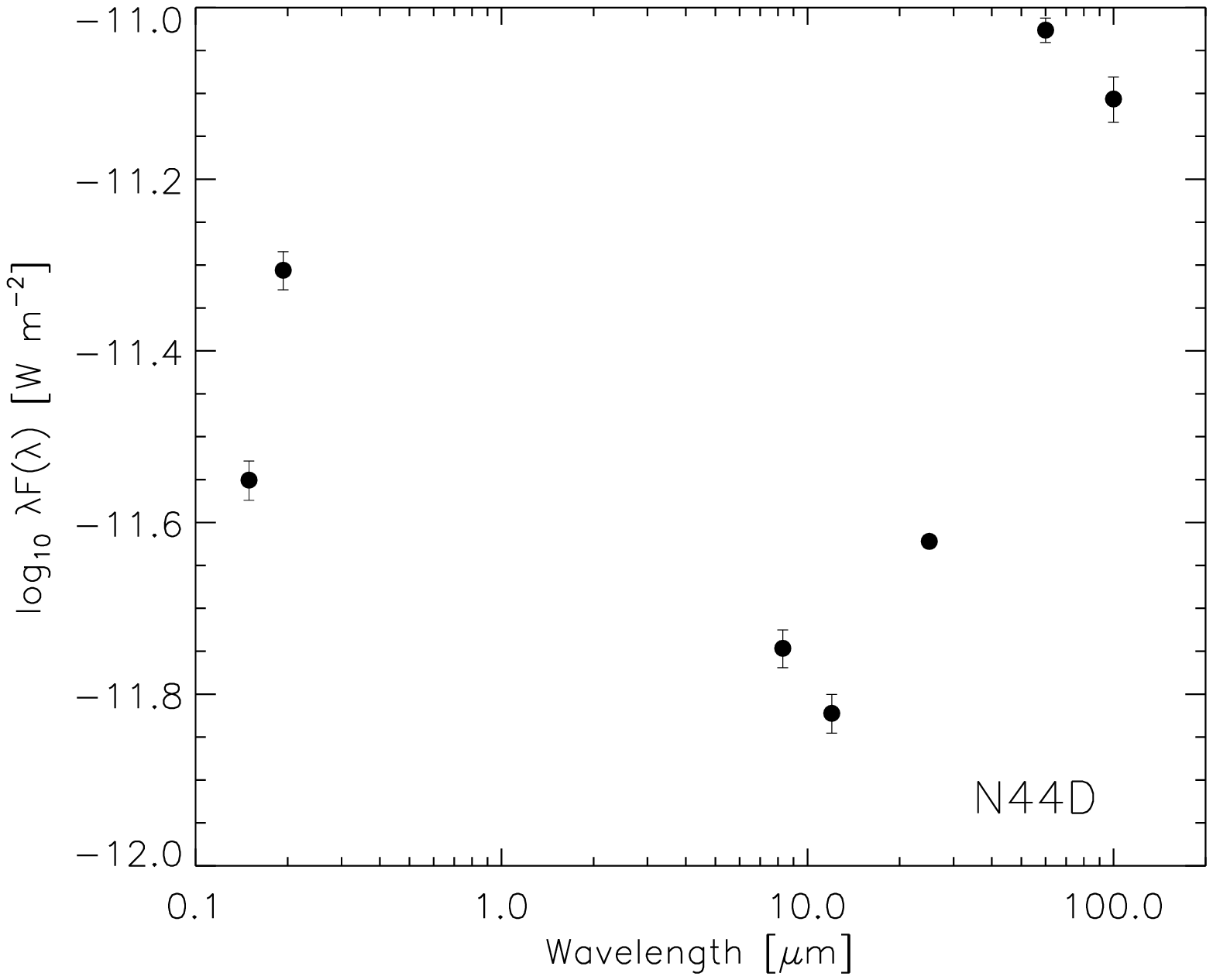}
\hfill
\epsfxsize=8.8cm
\epsfbox{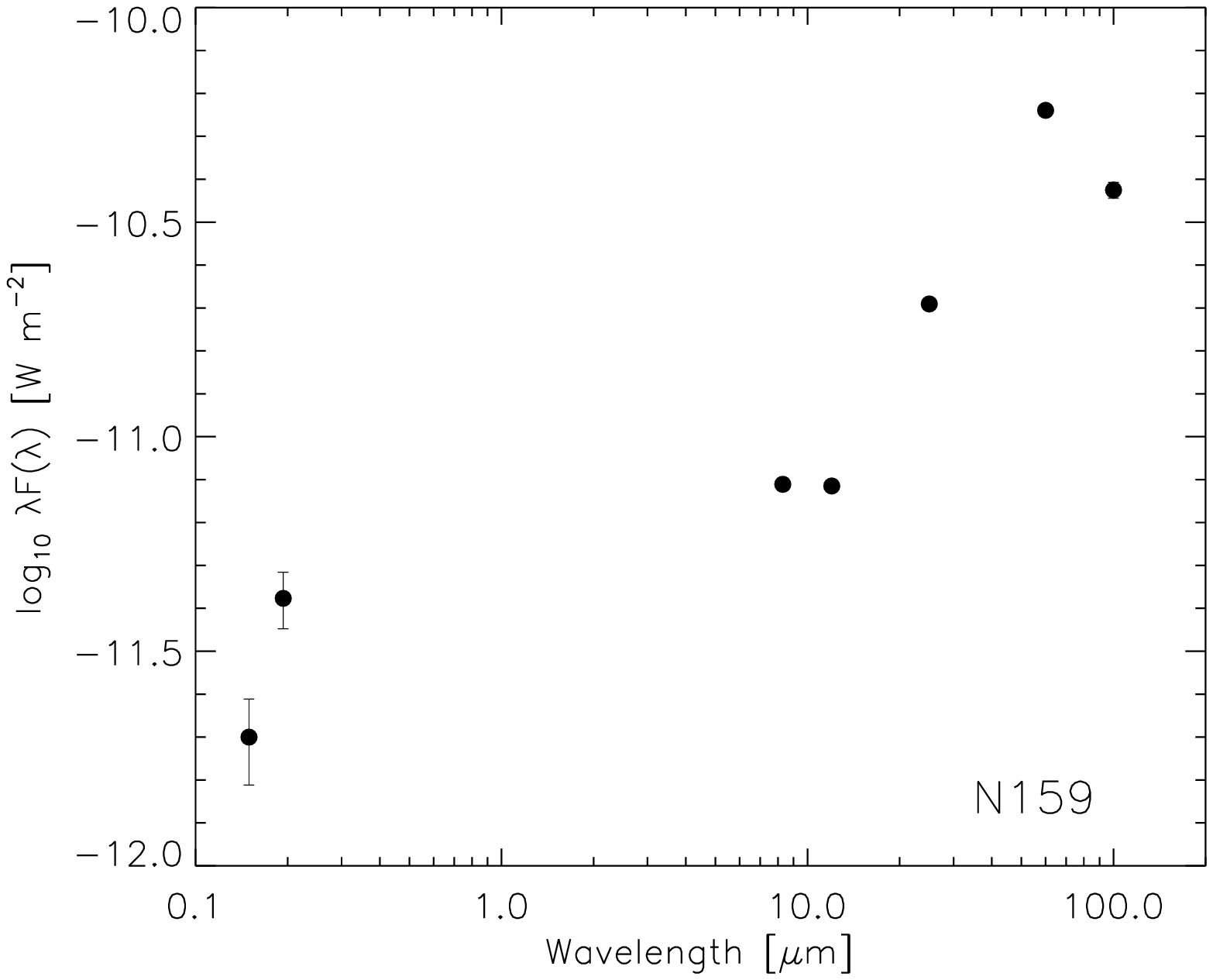}}
\caption{\label{fig:sed}
SEDs in 4.9{\arcmin} apertures for four \hii regions, sorted by extinction
(left to right; top to bottom):
N51D ($\sim$0 mag at \hans), N113 N ($\sim$0.15 mag at \hans), 
N44D ($\sim$0.25 mag at \hans), and N159 ($\sim$0.6 mag at \hans).
The \ha extinctions are determined using the Balmer decrement within
4.9{\arcmin} as measured by \protect\citet{cap85}, assuming a Milky 
Way extinction curve and a foreground screen dust distribution.
The logarithm of the energy flux $\lambda {\rm F}_{\lambda}$ in W\,m$^{-2}$
is shown at 1500{\AA}, 1900{\AA}, 8.3$\micron$, 12$\micron$, 
25$\micron$, 60$\micron$, and 100$\micron$.  The 8.3$\micron$ 
flux can be larger than the 12$\micron$ flux due to emission from 
the unidentified infrared bands, most plausibly associated with 
polycyclic aromatic hydrocarbons \protect\citep[e.g.][]{puget89}.  
These data are not corrected
for foreground galactic extinction, but are color-corrected
(see text for details).
Note that some error bars are smaller than the plotting symbols.
}
\end{figure*}

We selected a sample of 
\hii regions with measured Balmer decrements in 4.9{\arcmin} diameter apertures
from the sample of \citet{cap85,cap86}. We retained the use
of the 4.9{\arcmin} aperture to ensure a common aperture for all attenuations:
the adoption of a larger aperture (up to 20$\arcmin$, depending
on source size and morphology) gave very similar
attenuations at better than the 10\% level (and similar values of UV spectral
slope $\beta$ at better than the 0.5 level). 
The locations of the \hii regions are indicated in Fig.\ \ref{fig:image}.
An explicit selection was made for isolation: any \hii regions from 
\citet{cap85,cap86} with significant contamination 
at any wavelength (determined visually) from other, 
nearby \hii regions were omitted from the sample.

To correct aperture flux measurements for `sky' contamination, 
local sky levels are estimated using a centered annulus 
with an inner radius of 20{\arcmin} and an outer radius of
30{\arcmin}.
The sky annulus is split into octants, and the median level within each 
octant is determined.  In constructing the mean and RMS of the 
local sky, the highest three octants are discarded, to ensure against
contamination of the sky level by nearby emission.  In (rare) cases
where there is no nearby emission, this approach results in a slight systematic
underestimate of at most 1$\sigma$ in the sky level, or a 
corresponding overestimate of at most a few percent
in the \hii region flux. In most
cases, however, this method largely negates the effects of
emission in the sky annulus and reproduces manually 
estimated sky levels to within 1$\sigma$ with no hint of systematic 
offset.

Aperture corrections are typically not required for most of the 
data.  However, because the FWHM of a point source in the 8.55 GHz radio map 
is 210{\arcsec}, a modest aperture correction is
applied.  We derive this aperture correction
empirically by  smoothing the observed image with a 210{\arcsec} FWHM
Gaussian beam and comparing the aperture flux in the 
convolved and original images.  The ratio 
of these fluxes is adopted as the aperture correction.  
This procedure is tested by re-convolving the convolved
image (i.e. how well can one recover the flux in the original image
by aperture correcting the flux measured from the convolved image). 
This procedure appears to be accurate
to $\sim$5\%.  Typical aperture corrections are $\sim$15--25\%, depending 
on aperture size and source morphology.

Color correction in the \fuv and \fir is required (the \ha and
radio do not require color correction because of the narrow 
bandpasses).  
Fluxes are corrected to the nominal effective wavelength
of the filters.  \fuv color corrections, in terms of the observed
1500{\AA} to 1900{\AA} flux ratio, are calculated using a sequence
of {\it IUE} spectra that are artificially reddened using 
a SMC extinction curve (used because it lacks
a 2200{\AA} bump, leading to a more stable color correction).  
Using different {\it IUE} spectra and/or a LMC extinction
curve to artificially redden the spectra leads to differences in color
correction smaller than $\sim$3\%.
\fir color corrections
are estimated from the observed 60$\micron$ to 100$\micron$ 
flux ratio, using the multiplicative corrections from 
the {\it IRAS} Explanatory Supplement (1.09$\,\pm\,$0.01 at 60$\micron$ and
1.01$\,\pm\,$0.01 at 100$\micron$, for the 
range of observed \hii region \fir colors).  Color correction 
for the 8.3$\micron$, 12$\micron$, and 25$\micron$ fluxes
was not attempted due to the complexity of the \fir 
spectrum at those wavelengths and negligible contribution of the \fir 
emission from those wavelengths to the integrated \fir emission.

In Fig.\,\ref{fig:sed}, we show examples 
of our \hii region SEDs (omitting the radio flux
point because it makes a negligible contribution
to the total bolometric flux of a star formation region, and 
the \ha data point because it is line emission).  
The error bars (which are in some cases smaller than the plotted symbol) 
reflect sky level uncertainties: a 5\% uncertainty is added to the sky error
in the \fuv to account for the color correction (and adding
an arbitrary modest contribution for `flat fielding').  A 5\%
uncertainty is added to the sky error in the radio to account for aperture
correction errors.  
\ha errors include an arbitrary
5\% flat fielding and continuum subtraction uncertainty.
Calibration and image alignment uncertainties are not accounted for
in the error bars.
We do not correct for 
foreground galactic extinction in constructing the SEDs, but correct
for it when quantifying the \hii region attenuations.
We present the \hii region fluxes and uncertainties
within the 4.9{\arcmin} apertures in Table \ref{tab:flux}.
The fluxes in this Table {\it have not} been corrected for galactic
foreground extinction.

\section{The effects of dust in LMC \hii region SEDs} \label{sec:dust}

Fig.\,\ref{fig:sed} illustrates the significant effect of dust
on \hii region SEDs.  As expected, 
increasing amounts of dust progressively
depress the UV flux, re-processing this energy into the
\firns. By using our understanding of the physics of nebular
emission and the spectra of different stellar 
populations, we can produce quantitative attenuation estimates.
In this section, we apply the various
attenuation estimators described previously and compare the results.
For reference, our measurements of the
\hii region attenuation and UV spectral slope $\beta$ within 
the 4.9{\arcmin} apertures, corrected for foreground galactic 
extinction, are presented in Table \ref{tab:atten}.

\subsection{Comparing estimates of \ha attenuation}

\begin{figure}[tbh]
\epsfxsize=\linewidth
\epsfbox{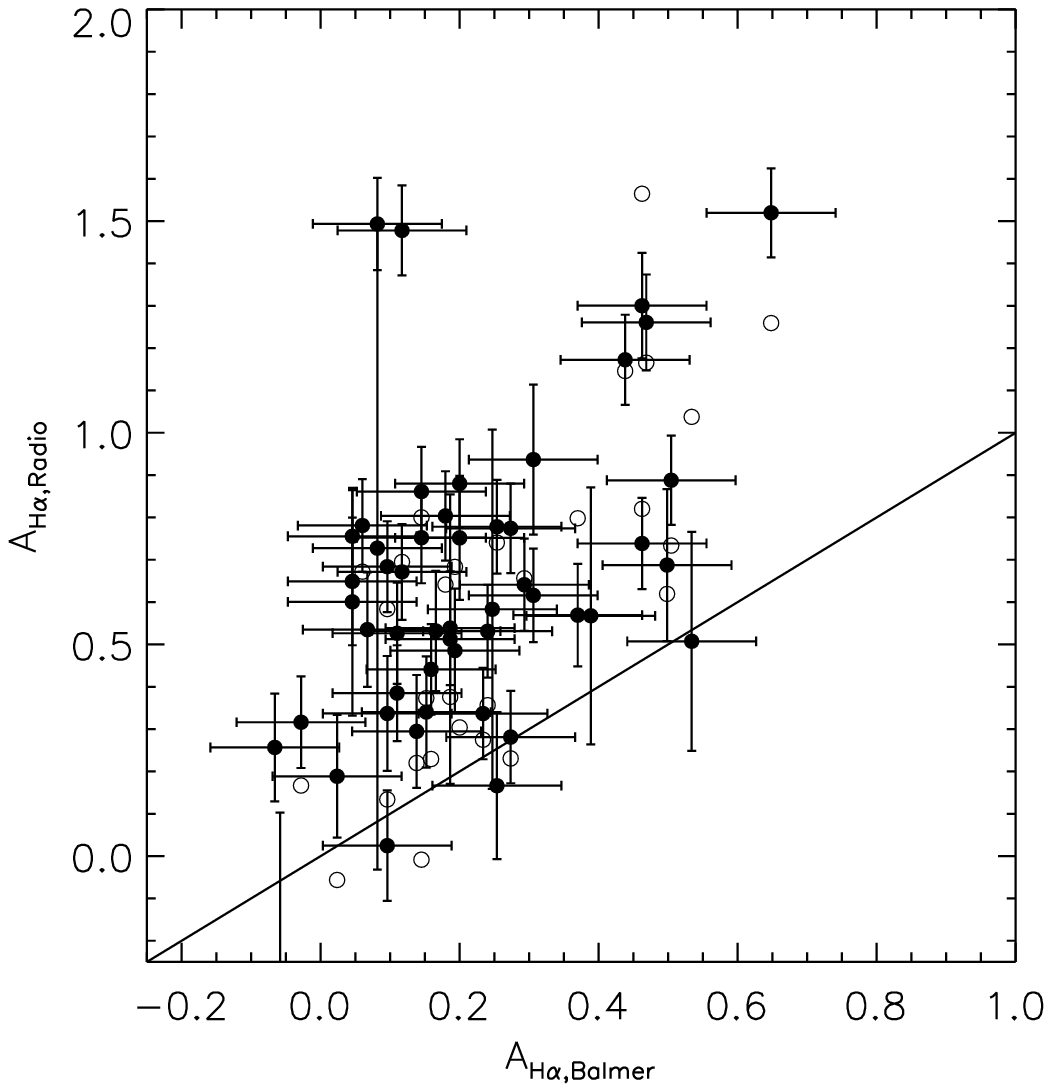}
\caption{\label{fig:cd86} Solid points denote radio-derived \ha attenuations 
(\aradns) against Balmer-derived attenuations (\abalmns), using
the 4.9{\arcmin} aperture of \protect\citet{cap85,cap86} for 
50 of our 52 \hii regions.
The Balmer-derived attenuation \abalm
is derived using \hahb values from \protect\citet{cap85}.
Open points denote
the radio-derived attenuations from \protect\citet{cap86}, 
for the 29 \hii regions in common with our clipped sample of 50 \hii 
regions.  The solid line denotes the relationship expected if
\abalm equals \aradns. }
\end{figure}

\citet{cap86} found that radio-derived \ha attenuations of LMC \hii 
regions, \aradns, exceeded the Balmer-derived \ha attenuations, \abalmns.
This result was confirmed using independent radio data for LMC \hii regions
\citep{ye98} and for \hii regions in \ M51 \citep{vdh88}.
This systematic offset was attributed to clumped dust, which 
makes the effective attenuation curve grayer (cf. panel {\it (c)} of 
Fig.\,\ref{fig:radtrans}).  

We re-examine this question using
\hahb values from \citet{cap85} corrected for Galactic extinction,
in conjunction with 
\arad values derived from our galactic extinction-corrected
aperture \ha fluxes and aperture-corrected radio fluxes.  
We discard N30BC (no Balmer decrement) and N158A (radio detection 
less than 2$\sigma$). Uncertainties in \arad come from propagating
the uncertainties in the SED measurements
and the adopted electron temperature.
The uncertainty in \abalm is calculated to be 9\% by propagating 
a 4\% uncertainty in \hahb 
\citep{cap85} and the uncertainty in the electron temperature.
We plot our values for \arad and \abalmns,
the line of equality for reference, 
and the original data from \citet{cap86} in Fig.\,\ref{fig:cd86}.
For 29 \hii regions in common we find a good overall agreement
between our values of \arad and those from \citet{cap86},
with a mean offset of 0.1 mag (ours are larger)
and RMS of 0.3 mag.  Omitting the 4 \hii regions with discrepancies
larger than 0.5 mag (three high values and one low value)
reduces the offset to 0.06 mag and the RMS to 0.14 mag. For 
reference, our median \arad formal uncertainty for these 29
\hii regions is 0.11 mag.
We confirm the offset between \arad and \abalm
observed by \citet{cap86}: we discuss the origin of this offset 
in \S\ref{subsec:prob}.

\subsection{Comparing estimates of far-UV attenuation} \label{compstar}

We now assess the consistency
of indicators of the attenuation suffered by the starlight from 
young stellar populations.  We compare reddening estimates, and
inter-compare these with estimates of the attenuation in the UV
derived using UV/\fir energy balance.  When comparing
these attenuation indicators, we also present the na\"{\i}ve expectation
from a screen model with LMC and SMC Bar-type dust.  The intention
of showing these models is to enable one to use 
existing intuition to help understand the results. 
These models do not represent the true situation. In particular, it 
would be incorrect to differentiate between possible
extinction curves on the basis of these data because
radiative transfer effects 
complicate the interpretation of these plots.

\subsubsection{Estimating the attenuations} \label{sec:estimate}

In this section, we discuss the construction of 
\fuv attenuation estimates using the \fuvns/\fir ratio and
outline the methods used to estimate the stellar $E(B-V)$
and the UV spectral slope $\beta$.

As discussed in \S \ref{sec:tut}, it is possible to estimate
\fuv attenuation using a \fuvns/\fir flux ratio.  This can 
be done using a simple energy balance.  However, we 
use the flux ratio method of \citet{fluxrat} to estimate the attenuation,
to allow us to fully take into account uncertainties due to dust
geometry, type and stellar populations.  
The 1500{\AA} attenuation derived
using this method, as described in \S \ref{sec:tut}, is
approximated to within 0.03 mag by eqn.\ \ref{eqn:fr}.

To apply this method
we construct an integrated \fir $8 - 1000$$\micron$ flux following 
\citet{fluxrat}.  The 8.3{\micron}, 12{\micron}, 
25{\micron}, 60{\micron} and 100{\micron} fluxes
were numerically integrated to provide an estimate of the 
$8 - 120${\micron} flux.  The 60{\micron} and 100{\micron} fluxes
were then used to define
a dust temperature for a $\beta = 1$ dust emissivity model to
extrapolate to 1000{\micron}.  This extrapolation, which adds 
typically $\sim$25\% to the flux, was tested against raw
178$\arcsec$ square aperture fluxes from the ISO archive.
To within a constant scaling that accounts
for aperture correction and pointing offsets, 
the fluxes and model extrapolations
are within 10\% of each other at 120{\micron}, 150{\micron}, and 
200{\micron}.  
We assume no Lyman continuum ionizing radiation radiation is absorbed
by dust internal to the \hii region: assuming that 50\%
of the ionizing radiation is absorbed by dust produces around
a $-0.15\sigma$ systematic offset in the estimated attenuation.  

Alternatively, a more `standard' manner to compute the attenuation
of starlight from the central young cluster utilizes the 
measurement of reddening based on the optical stellar color excess
$E(B-V)$. For 45 particular LMC \hii regions, \citet{cap86} provide
$E(B-V)$ values, which we correct for galactic foreground reddening. 
However, these values are
average values for a number of stars {\it near} the \hii region.
Because some of these stars may be well in the foreground or background
relative to the \hii region, the average reddening does not 
necessarily reflect the attenuation toward the central star
cluster in the \hii region.

Finally, one can estimate the attenuation of the central star cluster using
the UV spectral slope $\beta$,
as estimated using the 1500{\AA} $-$ 1900{\AA} color
and as defined in eqn.\ \ref{eqn:beta}.
Uncertainties in $\beta$ are derived from  the
random errors in the \fuv flux, and typically become larger for
more highly-reddened \hii regions (as they are typically more
attenuated and are therefore fainter).    
The $\beta$ values of 4 UV-bright star formation regions were
tested against 10{\arcsec}$\times$20{\arcsec} 
{\it International Ultraviolet Explorer} (IUE) UV spectral slopes.
We find that 3/4 of the IUE $\beta$ values
agreed with the Rocket UV-derived $\beta$ values 
to better than 0.5 in UV spectral slope.
This is a remarkable
agreement considering the huge differences in aperture size
(10{\arcsec}$\times$20{\arcsec} compared to a 4.9{\arcmin} diameter
aperture).  It is worth briefly noting that for these four \hii regions
with {\it IUE} spectra 
there is no evidence for a systematic difference between 
values of $\beta$ derived using a synthetic 1500{\AA} $-$ 1900{\AA} color
and using the wavelength windows defined by \citet{calzetti94}.

\subsubsection{Comparing the attenuations}

\begin{figure}[tbh]
\epsfxsize=\linewidth
\epsfbox{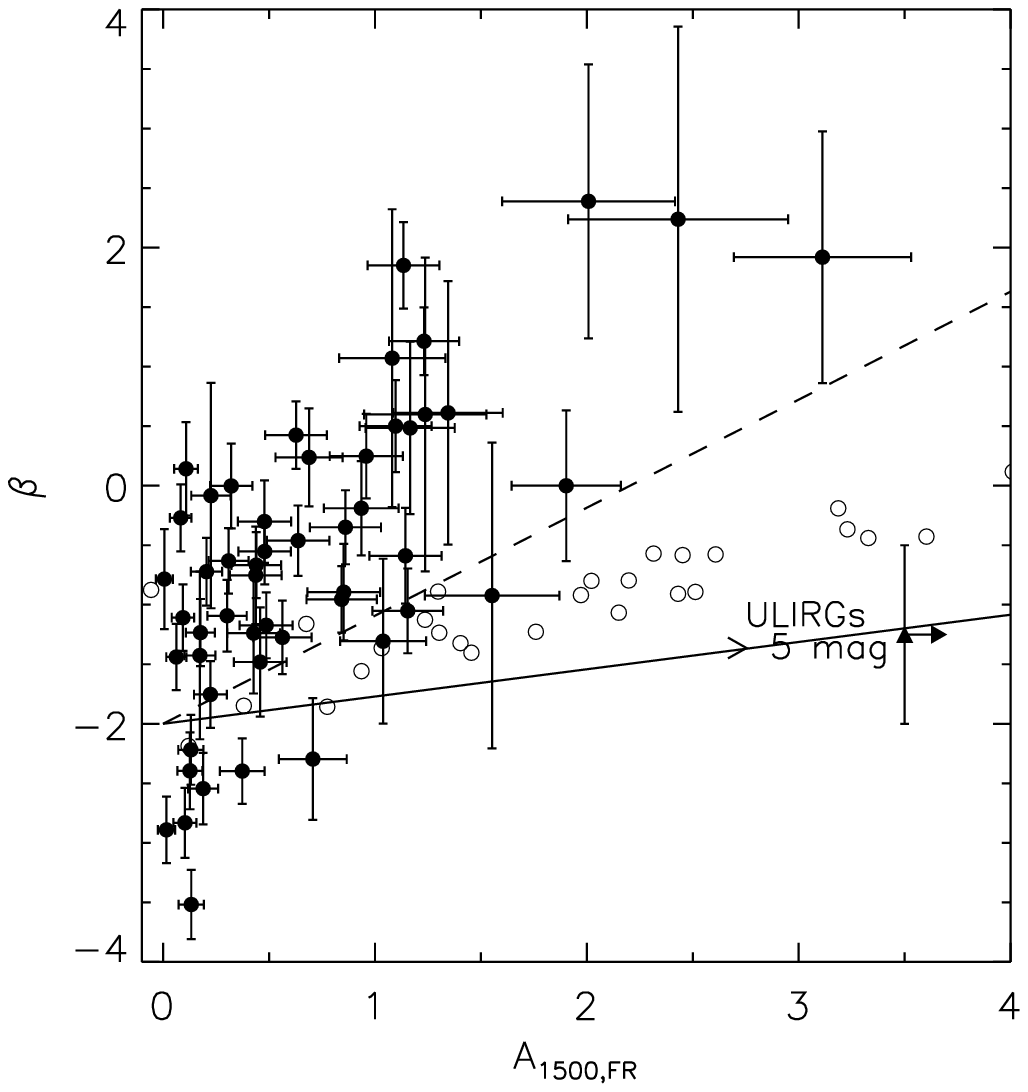}
\caption{\label{fig:beta} \fuv spectral slope $\beta$ against
the 1500{\AA} flux ratio-derived attenuation \afifns.  Quantities
were derived within the 4.9{\arcmin} aperture of 
\protect\citet{cap85,cap86} for our sample of 52 \hii regions.
The solid line denotes the expected correlation assuming 
a typical LMC extinction curve and the dashed line the SMC
extinction curve.  These lines, also shown in Figs.\ 
\ref{fig:optfr}--\ref{fig:betamod}, have been shown only to let the 
reader use their existing intuition to help understand the data;
however, it is inappropriate to try differentiate between different
extinction curves using these data because of the complicating
effects of radiative transfer (see e.g.\ Figs. \ref{fig:radtrans} and
\ref{fig:betamod}).
An intrinsic $\beta$ of $-2.0$ has been assumed.
Open points denote the starburst sample
of \protect\citet{calzetti94}: their values of $\beta$ were
calculated in a similar way to the LMC \hii regions.  Average
results from \protect\citet{meurer00} for a sample of 6 
ultra-luminous infrared galaxies are also shown, which
have estimated 1500{\AA} attenuations in excess of 5 magnitudes.  }
\end{figure}

We show the direct comparison between UV spectral slope $\beta$ and 
\afif in Fig.\,\ref{fig:beta}.  
Despite the significant scatter, and considerable
observational errors, there is a trend between 
$\beta$ and 1500{\AA} attenuation. 
Two points are apparent immediately.  

Firstly, the change
in \fuv spectral slope $\beta$ for a given 1500{\AA} attenuation is 
{\it much} larger than expected using an average LMC
extinction curve (solid line).  Furthermore, the correlation is much steeper
than the correlation between $\beta$ and \fuv attenuation derived 
for starbursting galaxies (open points)
by \citet{calzetti94} and \citet{meurer99}.
The steepness of the relationship between $\beta$ and \afif
is likely due to radiative transfer effects, and does not imply that
the dust in LMC \hii regions is SMC-type: this point is discussed in
detail in \S \ref{subsec:beta}.
The existence and steepness of the correlation are robust: omitting 
the three points with highest \afifns, or even one quarter of the
data points 
with highest \afifns, does not significantly change either the significance
or slope of the correlation.

The open points were calculated using synthetic
1500{\AA} and 1900{\AA} fluxes derived from {\it IUE} spectra, and
\fir luminosities calculated as above for the \hii regions using 
data from \citet{calzetti94,calzetti95}.  Typical errors in 
$\beta$ and \afif for the starburst galaxies are $\sim 0.1$ and
$\sim 0.15$ mag respectively, which is smaller
than the (modest) scatter shown by the starbursts.
Finally, we also show average results from \citet{meurer00} for a sample of 6 
ultra-luminous infrared galaxies that
have estimated 1500{\AA} attenuations in excess of 5 magnitudes.

This comparison of empirical results demonstrates directly that the correlation
between $\beta$ and \fuv attenuation is highly variable.  
Although \hii regions are arguably the basic unit of 
star formation, the behavior of the $\beta$--attenuation correlation
differs strongly between \hii regions and intensely star-forming galaxies.

Secondly, the
scatter in the \hii region $\beta$--attenuation
correlation is large. This 
scatter indicates that the \fuv spectral slope is not necessarily
an accurate indicator of \fuv attenuation, which is not surprising
because the relationship between
$\beta$ and \fuv attenuation is highly dependent on 
dust/star geometry and \fuv extinction curve shape
\citep[e.g.][see also Fig.\ \ref{fig:radtrans}]{clump2}.
We discuss Fig.\ \ref{fig:beta} in more detail in 
\S \ref{subsec:beta}, while investigating possible reasons 
for the steep slope and scatter and discussing
some implications of this result.

\begin{figure}[tbh]
\epsfxsize=\linewidth
\epsfbox{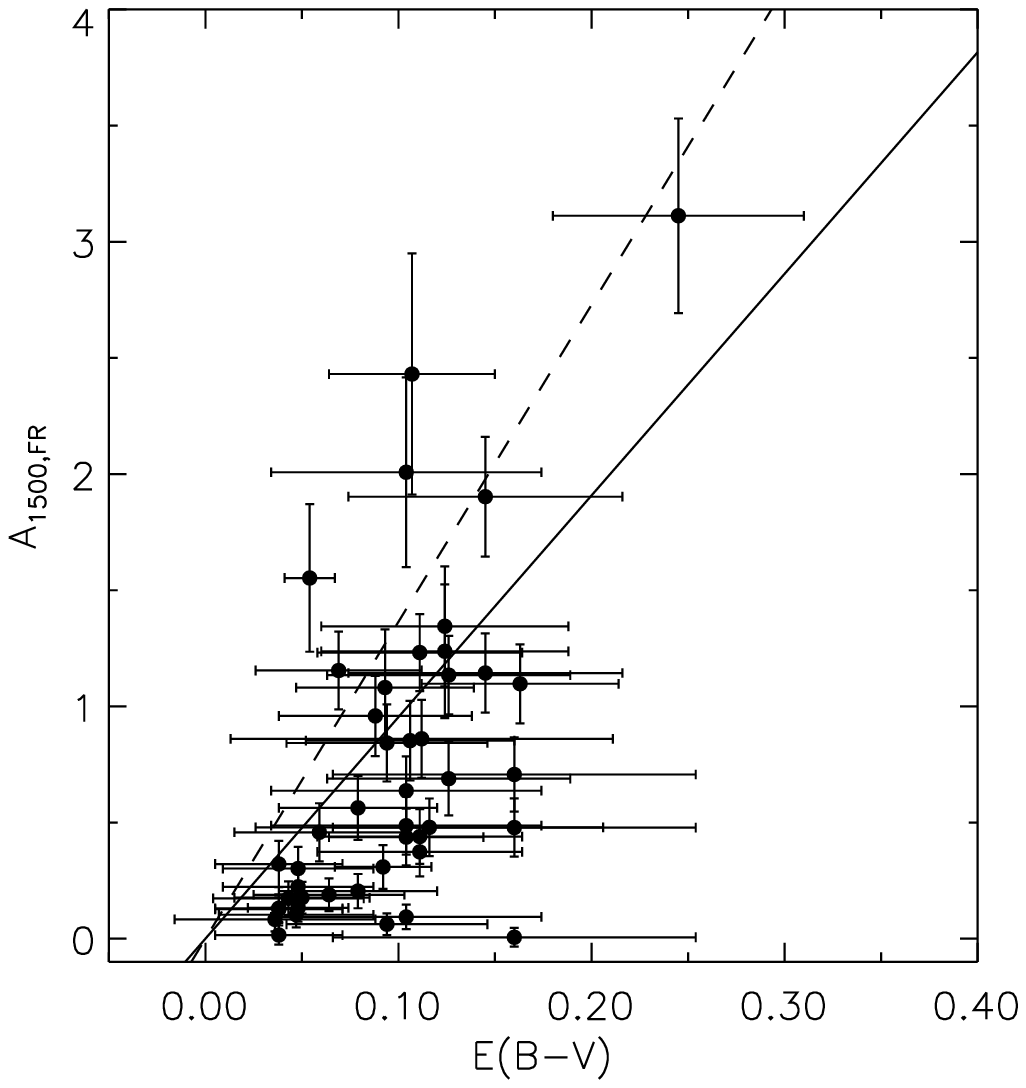}
\caption{\label{fig:optfr} Optical color excess $E(B-V)$ against
the 1500{\AA} flux ratio-derived attenuation \afifns.  The 
$E(B-V)$ values were derived for the overall region around 
the 45 \hii regions of interest.  Flux ratio-derived attenuations
were derived within a 4.9{\arcmin} aperture.
The solid line denotes the expected correlation assuming 
a typical LMC extinction curve and the dashed line the SMC
extinction curve. }
\end{figure}

\begin{figure}[tbh]
\epsfxsize=\linewidth
\epsfbox{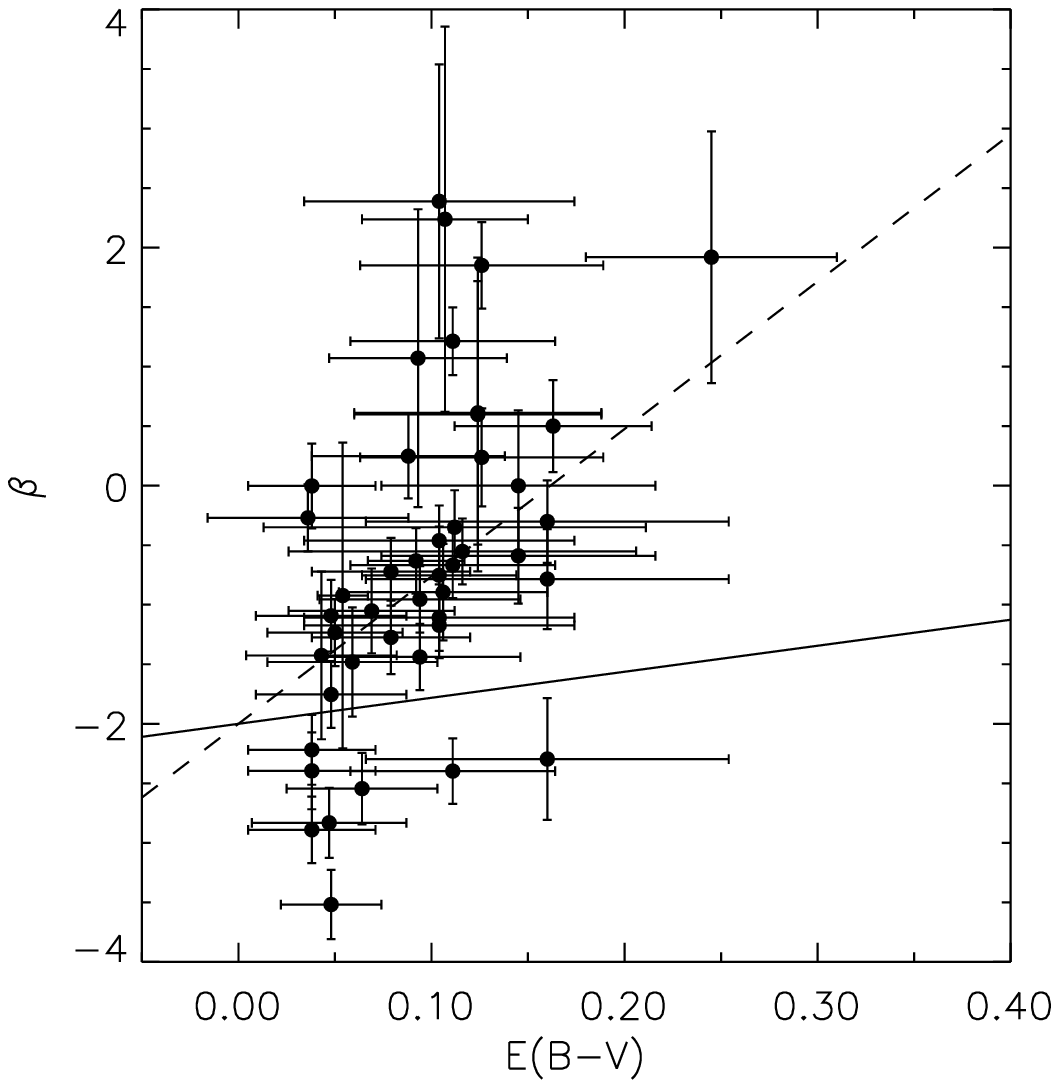}
\caption{\label{fig:optbeta} Optical color excess $E(B-V)$ against
the UV spectral slope $\beta$ derived within 4.9{\arcmin} apertures.  
Solid and dashed lines denote the expectations of a screen
model with LMC-type and SMC Bar-type dust respectively.  }
\end{figure}

\fuv attenuation may be 
estimated from $E(B-V)$, in conjunction with 
assumptions regarding the extinction curve and dust geometry.
In Fig.\,\ref{fig:optfr}, we show \afif against $E(B-V)$.    
In Fig.\,\ref{fig:optbeta}, we show the UV spectral slope $\beta$ against
$E(B-V)$.  Overplotted are expectations for the 
correlations assuming an average LMC extinction curve (solid line)
and SMC bar extinction curve (dashed line).
Despite the large scatters in both correlations, 
the correlations are significant at greater than the 99.9\% level, with 
Spearman rank correlation coefficients of 0.57 (\afifns--$E(B-V)$)
and 0.54 ($\beta$--$E(B-V)$).

From this discussion, we conclude that 
the values of \afifns, $\beta$ and $E(B-V)$ for our sample of \hii regions
are broadly consistent: at the very least,
as one increases, the other indicators tend to increase also.
The large scatter in the correlation, however, demonstrates that
the assumptions behind each of the methods are continually being
violated to varying degrees.

\subsection{Comparing \ha and far-UV attenuation} \label{compgasstar}

\begin{figure}[tbh]
\epsfxsize=\linewidth
\epsfbox{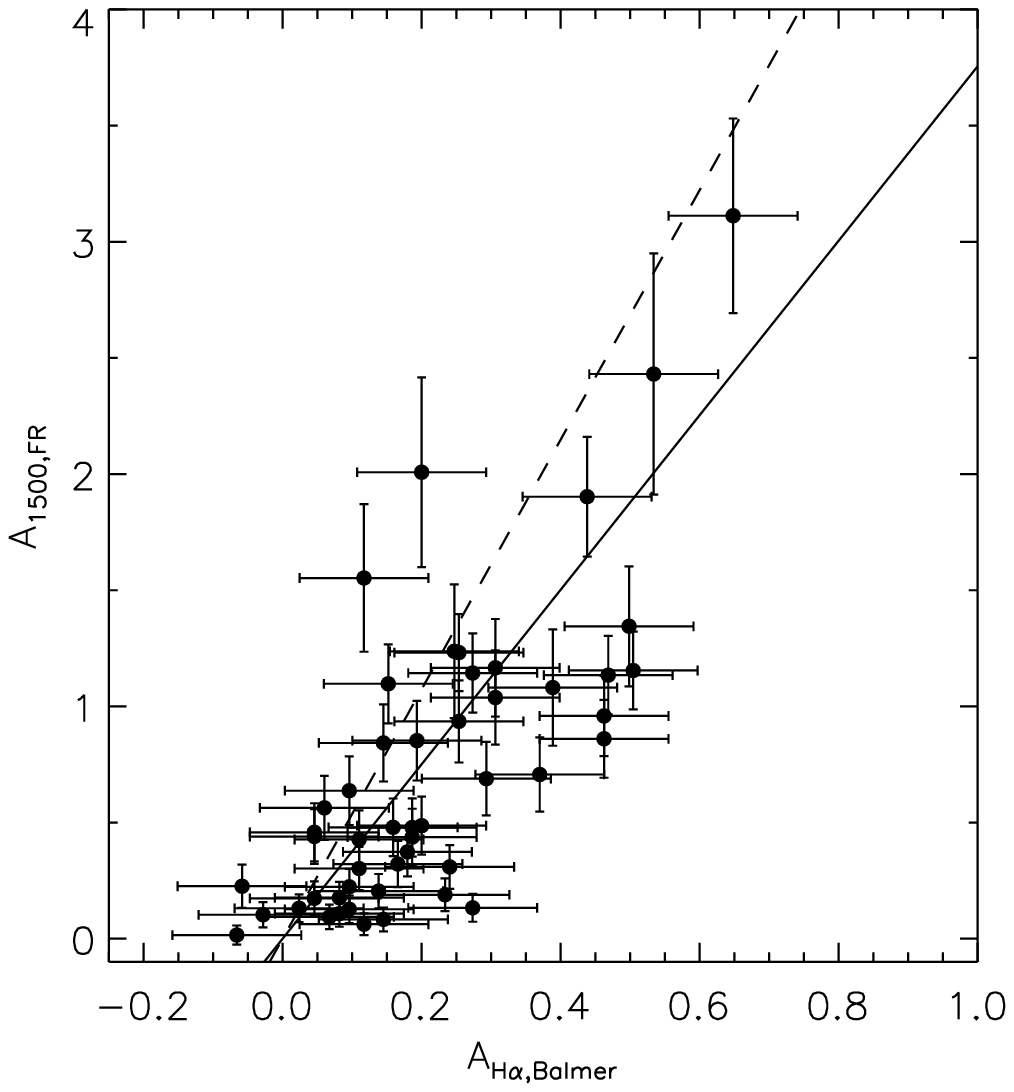}
\caption{\label{fig:frbalm} The 1500{\AA} flux ratio-derived attenuation
\afif against the Balmer-derived \ha attenuation \abalmns.
Attenuations were derived within the 
4.9{\arcmin} aperture of \protect\citet{cap85,cap86} for 
50 of our 52 \hii regions. 
Solid and dashed lines denote the expectations of a screen
model with LMC-type and SMC Bar-type dust respectively.  }
\end{figure}

\begin{figure}[tbh]
\epsfxsize=\linewidth
\epsfbox{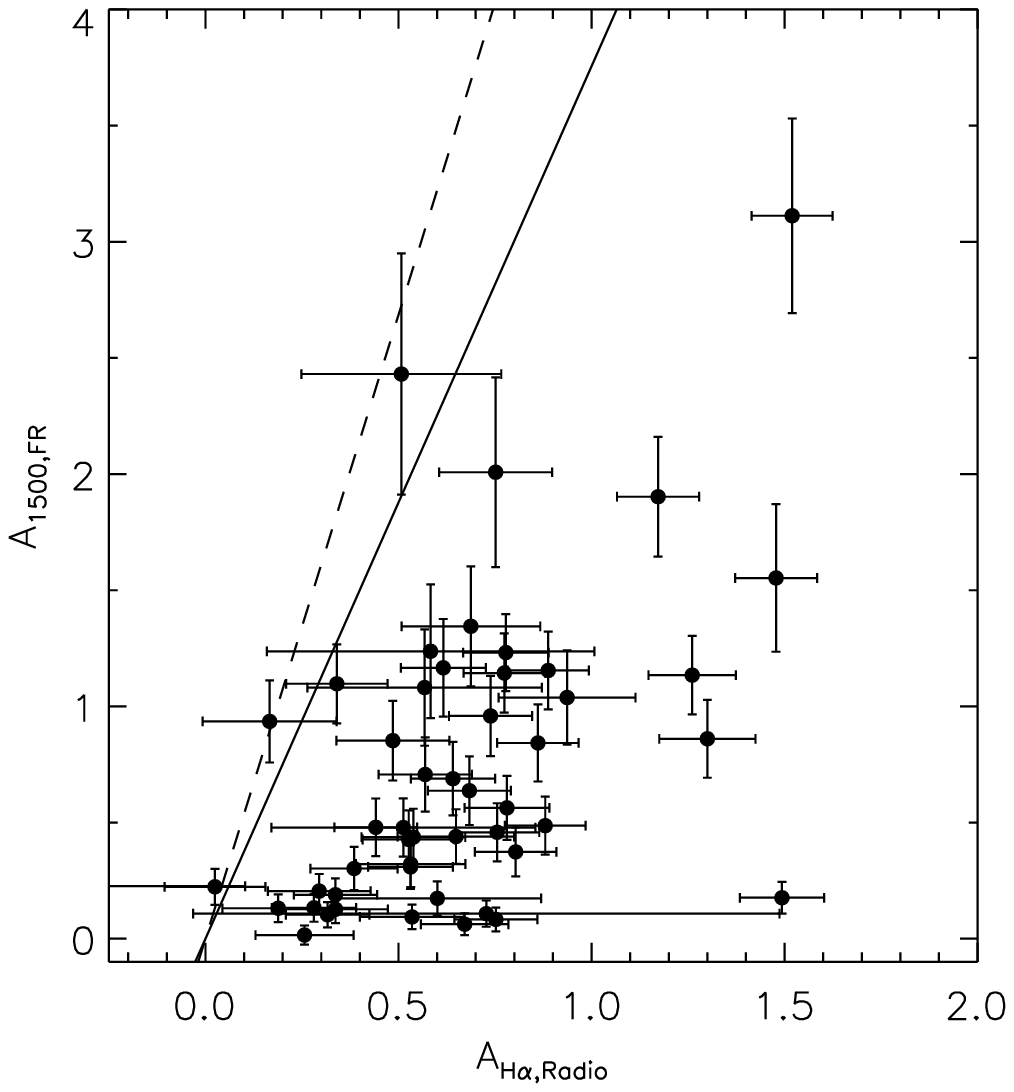}
\caption{\label{fig:frrad} The 1500{\AA} flux ratio-derived attenuation
\afif against the radio-derived \ha attenuation \aradns.
Attenuations were derived within the 
4.9{\arcmin} aperture of \protect\citet{cap85,cap86} for 
50 of our 52 \hii regions.  
Solid and dashed lines denote the expectations of a screen
model with LMC-type and SMC Bar-type dust respectively.  }
\end{figure}

We have tested \ha and \fuv attenuation estimates independently
for internal consistency and now turn to cross-comparing them.
\citet{cap86} found
a highly scattered correlation between $E(B-V)$ to \abalmns.
We explore other correlations between 
\ha and \fuv attenuation indicators in Figs.\,\ref{fig:frbalm} and
\ref{fig:frrad}, where we show the flux ratio-derived 1500{\AA}
attenuation, \afif, against the Balmer-derived \ha attenuation,
\abalmns, and the radio-derived \ha attenuation, \aradns.

In Fig.\,\ref{fig:frbalm} we plot
\afif against \abalmns.
Although there is considerable scatter, the
overall correspondence between \afif and \abalm (significance $>$ 99.9\%; 
Spearman rank correlation coefficient of 0.72) is strong. The error-weighted
mean ratio of \afifns/\abalm is 3.4$\,\pm\,$0.3 with a RMS of 
1.8 for 50 \hii regions.  This correspondence is as
expected after accounting for  radiative transfer effects (cf. panel {\it (c)}
of Fig.\,\ref{fig:radtrans}).

In Fig.\,\ref{fig:frrad} we show the correlation between \afif 
and \arad
(significance $>$ 99.9\%; 
Spearman rank correlation coefficient of 0.47).
The error-weighted
mean ratio of \afifns/\arad is 1.3$\,\pm\,$0.1 with a RMS of 
0.7 for 50 \hii regions. This relationship is more complex than
the previous one between \afif and \abalmns.
For example, there are a considerable
number of \hii regions with either much lower or higher \arad
than would be expected on the basis of the measured \afifns.
This extra scatter will be discussed in more detail below. 
Despite some oddities, there is a basic correspondence between 
\ha and \fuv attenuations in our sample of LMC \hii regions, 
lending credibility to both sets of attenuation indicators.

\section{Discussion} \label{sec:disc}

The observed values of \abalmns,
\aradns, \afifns, $\beta$ and $E(B-V)$ are generally consistent,
especially 
given the uncertainties and the range of behaviors expected from
radiative transfer effects.
However, there are two discrepancies 
that deserve further discussion: the highly scattered correlation 
between \abalm and \aradns, and the deviant behavior of the 
highly scattered correlation between \afif and $\beta$. 

\subsection{Are there problems with radio-based \ha attenuations?}
	\label{subsec:prob}

\begin{figure}[tbh]
\epsfxsize=\linewidth
\epsfbox{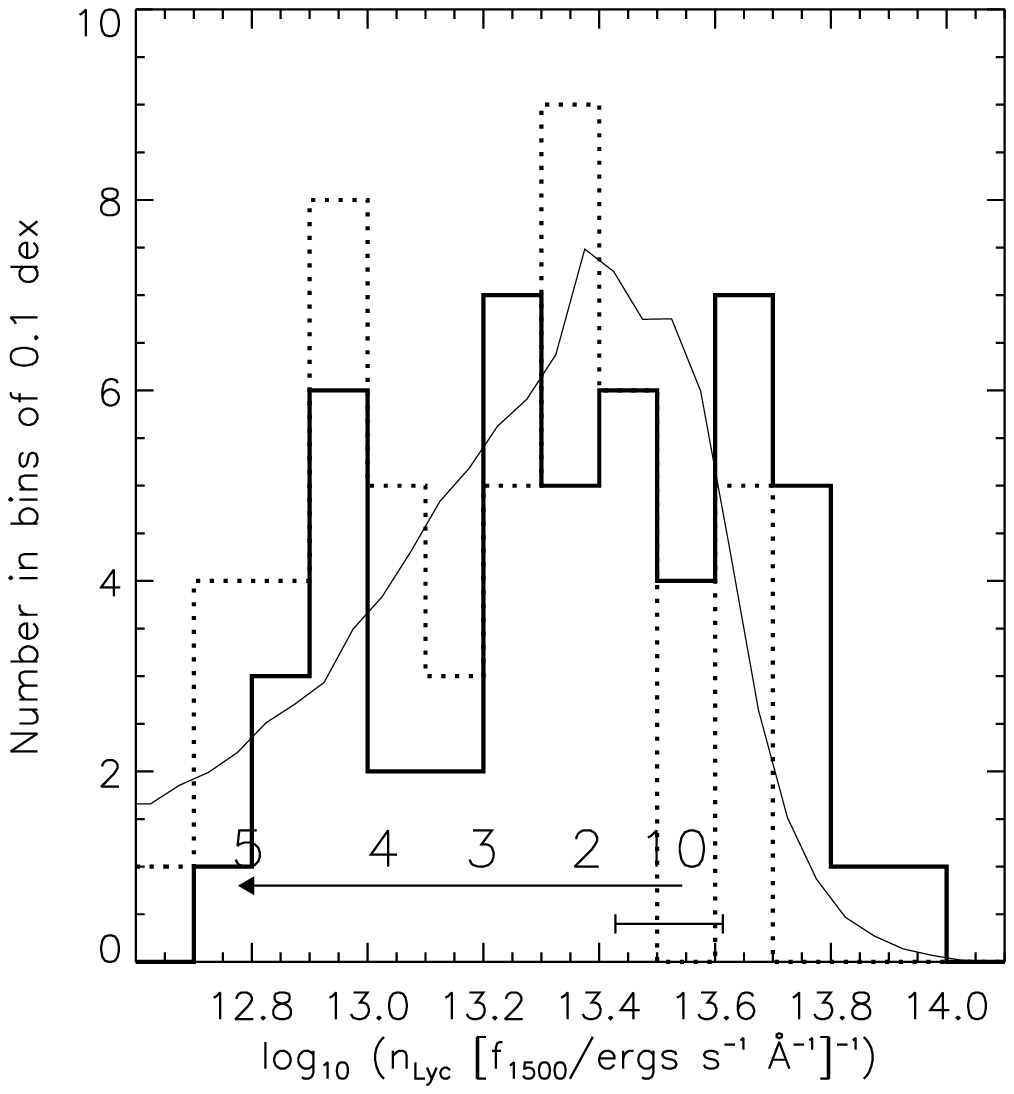}
\caption{\label{fig:hist} Distribution of the number of 
ionizing \hii region photons, per unit 1500{\AA} flux.  The number 
of Lyman continuum photons required to produce the observed 
\ha flux (assuming Case B recombination) has been corrected
assuming radio-derived \ha attenuations (thick solid line) and 
Balmer-derived \ha attenuations (thick dotted line).  The 1500{\AA}
flux has been corrected using the flux ratio method (the 
flux ratio method was earlier demonstrated to give attenuations
which were not inconsistent with other attenuation indicators).  The predicted
ratio $n_{\rm Lyc}/f_{1500}$ is shown for a single burst \peg model with 
a range of ages in Myr along the arrow at the bottom. This model
assumes 1/3 solar metallicity: models with metallicities within a factor
of three of this value give maximum ratios less than 0.05 dex different
from the 1/3 solar model.  We also show the effect of 
slope changes of $\pm$0.5 from the Salpeter value for the zero
age point (error bar).  Also shown is a model curve for $\sim10^5$ 
simulated \hii regions with ages between and 0 and 6 Myr, and total
stellar masses between 500${\rm M_{\sun}}$ and 8000${\rm M_{\sun}}$.
Scatter from small number statistics at the upper end of the IMF
and observational errors was accounted for 
(the curve for radio-derived errors is shown;
the curve for Balmer-derived errors is essentially identical).
Fluxes were derived within a 4.9{\arcmin} aperture.  }
\end{figure}

The systematic discrepancy between \arad and \abalm is well-known and
generally attributed to a relatively gray
effective attenuation curve, which is due to radiative transfer effects
\citep[e.g.][]{cap86,vdh88,ye98}.  We confirm the empirical result,
but suggest that the interpretation should be examined in more detail. 
A comparison of Figs.\,\ref{fig:frbalm} and \ref{fig:frrad} shows that 
the 1500{\AA} attenuation correlates with considerable scatter
with \abalmns, but correlates with a much larger scatter
with \aradns.  This behavior
suggests a limitation in the values of \aradns, 
rather than a limitation with \abalmns. This runs contrary to
the typical argument that \arad is robust because it
is oblivious to radiative transfer effects.

We are in a position to explore the relationship between
\abalm and \arad further than existing studies 
because of our expanded wavelength coverage.  Previous studies often involved
only the Balmer-derived and radio-derived attenuations 
\citep[e.g.][]{cap86,vdh88,ye98}, leaving the situation 
ambiguous. Other data (such as the UV and \fir fluxes) can be
used, in principle, to shed new light on the problem.

One sensitive diagnostic is the \hans-to-1500{\AA} flux ratio.  
This ratio is a sensitive indicator of the temperature of the
illuminating stellar population. Assuming Case B recombination, we
convert this ratio into $n_{\rm Lyc}/f_{1500}$, assuming 
0.45 \ha photons per Lyman continuum photon 
\citep{hummer87}: we plot this in Fig.\ \ref{fig:hist}.
Before converting the \ha luminosities into the number of Lyman continuum 
photons, we correct for \ha attenuation using either radio-derived
(thick solid line) or Balmer-derived (thick dotted line) attenuations.
The fraction of Lyman continuum photons absorbed by 
dust before being down-converted to \ha and radio 
is not accounted for. The loss of Lyman continuum photons 
to dust prior to their conversion to Balmer photons would 
simply increase the estimated $n_{\rm Lyc}/f_{1500}$: for 
reference, \citet{de92} estimates
the fraction of Lyman continuum photons lost to dust
at between 0.2 and 0.5 for six LMC \hii regions (corresponding to 
an underestimate of $n_{\rm Lyc}/f_{1500}$ of between 0.1 and 0.3 dex).
The 1500{\AA} flux is corrected for attenuation using
the flux ratio-derived attenuations (as is apparent from Fig.\ \ref{fig:beta},
attenuations estimated from $\beta$ and the calibration
of \citet{calzetti94} would be implausibly large and are
therefore not used here). 

The radio-derived attenuations lead to larger $n_{\rm Lyc}/f_{1500}$
than the Balmer-derived attenuations.  To be able to interpret the
differences between the two sets of values, 
we show the expected range of $n_{\rm Lyc}/f_{1500}$
from the \peg stellar population synthesis model (arrow; the numbers denote
\hii region age in Myr). 
The illustrated model has a metallicity of 1/3 solar.
Changing the metallicity by a factor of three in either
direction, or adopting the {\sc starburst99} model
of \citet{sb99} results in systematic offsets of
less than 0.05 dex.  
Changing the slope of the initial mass function
from the Salpeter logarithmic slope, $x=-1.35$, by $\pm$0.5 results
in changes of $\mp$0.1 dex in $n_{\rm Lyc}/f_{1500}$ at zero age
(the error bar in Fig.\,\ref{fig:hist}).
Comparison of the model predictions of 
$n_{\rm Lyc}/f_{1500}$ with the observations suggests that 
the radio-derived \arad may be overcorrecting the \ha luminosity, 
while the Balmer-derived \abalm may be more appropriate.  

To test the significance of this suggestion, we construct
a simple model.  The evolution of stellar populations in 
narrow mass ranges was calculated using the \peg stellar population 
model.  Around $10^5$ randomly-generated stellar 
populations with ages between 0 and 6 Myr and 
total masses between 500${\rm M_{\sun}}$ and 8000${\rm M_{\sun}}$ 
were extracted from a Salpeter IMF.  Model values of 
$n_{\rm Lyc}/f_{1500}$ were convolved with the observational
errors.  The model curve is shown in Fig.\ \ref{fig:hist}.
There is a significant tail of model $\log_{10} (n_{\rm Lyc}/f_{1500})$
values above the zero-age value of 13.55, which are attributable in 
part to the effects of observational error, and in part to 
small number statistics in populating the IMF of 
\hii regions with massive stars. We find that the 
observed distribution corrected with
radio-derived \ha attenuations has a number of rather large
values of $n_{\rm Lyc}/f_{1500}$, but it is not
inconsistent with model distributions.
The distribution corrected with
Balmer-derived \ha attenuations is also consistent with model
distributions.

So, while there is a definite discrepancy between observed
values of \abalm and \aradns, it is not possible at this stage
to identify unambiguously the limitations of each attenuation
indicator.  Folding in \fuv and \fir 
data can, in principle, give further insight, and, indeed, there
is a hint of limitations in the 
radio-derived \ha attenuations.  However, even given this extra data, it
is impossible to rule out that the distribution of 
$n_{\rm Lyc}/f_{1500}$, derived using radio-based \ha 
attenuations, is consistent with plausible stellar populations. 
More detailed investigations,
involving spatially-resolved maps of \hii regions at \hans, \hbns, and a 
number of radio frequencies, are 
required to fully address this question.

\subsection{The relationship between $\beta$ and \afif} 
  \label{subsec:beta}

In \S \ref{compstar}, we discussed the highly scattered and 
rather steep correlation between the UV spectral slope 
$\beta$ and the 1500{\AA} attenuation \afifns.  Both the slope
and scatter of the correlation run counter to (at least) observational
expectation: the correlation between $\beta$ and \afif for
starbursting galaxies is tighter and shows a 
shallower slope.  We
discuss possible interpretations of the steepness and scatter
in this correlation, and some implications of
this result for studies of the UV properties of high-redshift galaxies.

\begin{figure}[tbh]
\epsfxsize=\linewidth
\epsfbox{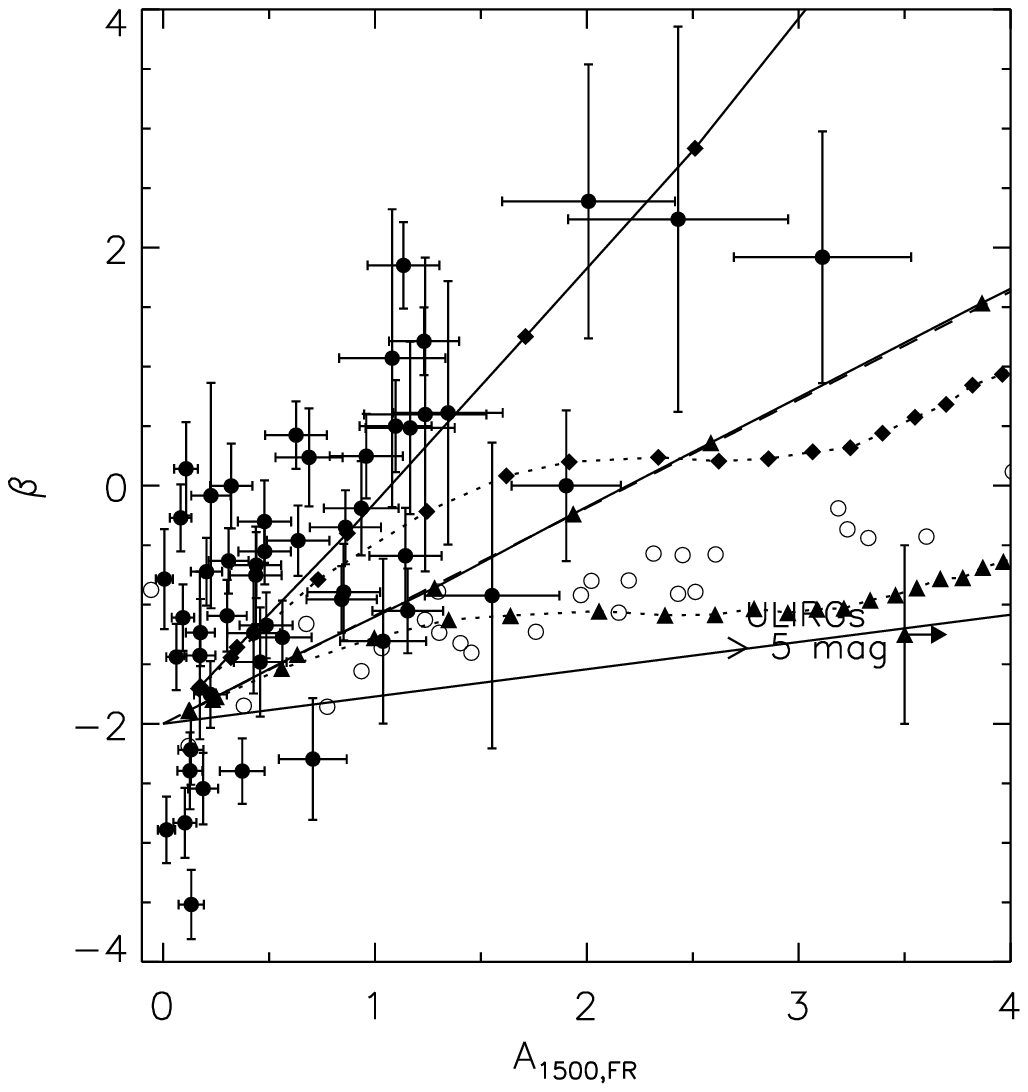}
\caption{\label{fig:betamod} \fuv spectral slope $\beta$ against
the 1500{\AA} flux ratio-derived attenuation \afifns.  
Symbols are as in Fig.\ \ref{fig:beta}.
The solid line without symbols denotes the expected correlation assuming 
a typical LMC extinction curve.  The dashed line (which is 
hidden under the solid line with triangles) denotes the SMC
extinction curve.  An intrinsic $\beta$ of $-2.0$ has been assumed.
These lines have been shown only to let the 
reader use their existing intuition to help understand the data;
however, it is inappropriate to try differentiate between different
extinction curves using these data because of the complicating
effects of radiative transfer.
To illustrate this point, radiative transfer model lines for LMC2-type
dust (for a supershell near 30 Dor; triangles) and 
SMC bar-type dust (diamonds) with 
homogeneous (solid lines with symbols) and clumpy (dotted lines
with symbols) geometries are overplotted. }
\end{figure}

We first address the possibility 
that the steepness and scatter of this correlation reflect
the data quality (or, for example, variations in Galactic 
foreground extinction), rather than 
the intrinsic properties of the \hii regions.
The quality of 
the data was discussed 
extensively in \S\S \ref{subsec:data:images} and \ref{compstar}.
Briefly, the zeropoint uncertainties in the
UV and \fir data used to construct the \afif estimates
are accurate to better than 10\%.
We have taken care to quantify and estimate appropriate sky
level errors (\S \ref{subsec:data:seds}), which are propagated into 
the final attenuation and UV spectral slope estimates.  
Furthermore, although the UV spectral slopes are difficult to check
against external comparison data, we found an agreement of better than
0.5 in slope in 3/4 test cases (comparing values derived from
10$\arcsec \times 20 \arcsec$ apertures 
with our values derived using a 4.9$\arcmin$ aperture).  We 
conclude that our error bars are fair, and that data quality issues
are not likely to cause the large scatter and steep slope we see in 
Fig.\ \ref{fig:beta} (re-plotted as Fig.\ \ref{fig:betamod}).  
Variations in Galactic foreground extinction
corresponding to $\Delta E(B-V)$ of $\pm 0.06$ mag 
produce a negligible spectral slope change
and variations in \afif of $\pm 0.5$ mag.  This kind of
variation is possible: \citet{oestreicher95}
discussed the evidence for variable foreground extinction in the direction 
of the LMC, but with insufficient spatial resolution 
for our purposes.  Given this, we
conclude that variable foreground extinction can contribute
significantly to the scatter, but cannot account for it in 
its entirety.

We turn to intrinsic properties of the \hii regions to explain 
the observation.
No more than 0.05 dex of scatter in 
$\beta$ and 0.15 mag of scatter in \afif can be generated by 
appealing to small number statistics at the upper
end of the \hii region IMF: this scatter is much smaller than 
the observed scatter. 
A promising possibility is the combined 
influence of differing extinction curves and dust/star geometry:
in \S \ref{sec:tut} we demonstrated that these can
strongly affect the relationship between \afif and $\beta$.  We 
demonstrate this more quantitatively in Fig.\ \ref{fig:betamod}, where
we replot Fig.\ \ref{fig:beta}, now including models from 
the {\sc dirty} radiative transfer model.  
These models, in the spirit of the earlier inclusion of LMC and SMC
screen models in earlier plots, are not meant to suggest that we
are trying to differentiate between different geometries or
extinction curves: rather, we are attempting to guide our intuition 
as to what effects can be plausibly caused
by extinction curve and geometry variations.

Variations in \hii region extinction curves and/or
star/dust geometries can produce a steep slope {\it and} a large
scatter in the $\beta$--\afif plane.  Slopes in this plot steeper
than those produced by a simple screen are possible because
of scattering (compare the SMC Bar-type extinction curve screen and shell 
model). However,
deviations from a perfect shell introduce significant scatter
and flatten the relation somewhat (the clumpy shell models have shallower
slopes than the homogeneous shell models). Lastly, it may
be possible to produce redder spectral slopes
and very low \afif values with either 
Galactic foreground extinction
variations or small amounts of dust directly along 
the line of sight, which produce significant reddening but relatively 
little \fir luminosity.  

We conclude that modest (and expected) variations in geometry can produce
the steep slope of the $\beta$--\afif
correlation and at least some of the scatter. 
This result carries with it serious implications 
for the use of $\beta$ as an accurate indicator of \fuv attenuation.
Neither ultra-luminous infrared galaxies nor our sample of LMC \hii regions
follow the relatively tight correlation between $\beta$ and \afif
established for starbursting galaxies.   
However, this correlation is utilized extensively for the estimation
of the SFRs of high-redshift galaxies  \citep[see e.g.][
for extensive testing of this correlation, and references therein]{adel00}.
Although the high-redshift star-forming galaxies are likely to be 
more similar to the low-redshift starburst sample of \citet{calzetti94}
than to our sample of LMC \hii regions, it is nevertheless a 
concern that the basic units of star formation (\hii regions) have
a drastically different behavior from star-bursting galaxies.
Clearly, much caution is warranted before 
estimating \fuv attenuations based on UV spectral slopes.

\section{Summary} \label{sec:conc}

We have used \fuvns, \hans, \firns, and 8.55 GHz 
radio images of the LMC to investigate the SEDs and attenuations 
of \hii regions drawn from 
the catalogs of \citet{cap85,cap86}.  We briefly
discussed the relative morphologies of the LMC in the \fuvns, optical
continuum, \hans, and \firns. 
In particular, the \fuv and \ha distributions within the LMC are 
different, implying that our findings regarding the effects of 
dust in the UV in LMC \hii regions may not apply to 
the LMC as a whole.

We constructed SEDs for the sample of \hii regions.  The effects
of attenuation on the SEDs are clear: \hii regions with 
larger Balmer decrements show a larger \fir to UV luminosity ratio,
and typically redder UV spectral slopes (Fig.\ \ref{fig:sed}).
To quantify these trends, we 
constructed attenuation indicators of both the ionized \hii region
gas (the Balmer decrement and radio to \ha luminosity ratio) and the
stars (the UV spectral slope $\beta$, optical stellar reddening, and 
the balance of \fuv and \fir luminosity).  Our principal findings
are as follows.

\noindent $\bullet$
Estimates of \ha and \fuv attenuation all correlate,
although often with substantial scatter (Figs.\ 
\ref{fig:cd86}--\ref{fig:frrad}).  Some of this
scatter is due to errors in the (often quite modest) attenuations.
However, much of this scatter is
naturally interpreted in terms of radiative transfer in 
non-trivial dust/star geometries.  

\noindent $\bullet$
The modest Balmer-derived \ha attenuations are systematically lower
than radio-derived \ha attenuations.  Previous studies have 
attributed this systematic offset 
to variations in the extinction within the beam.
In principle, more insight can be gained by incorporating
\fuv and \fir fluxes.  However, the inclusion of these data cannot, 
at this stage, unambiguously identify the source of the 
discrepancy between Balmer-derived and radio-derived \ha attenuations.
More work, involving spatially-resolved maps of \hans, \hbns,
and radio emission at a range of frequencies, is required to 
properly investigate this issue.

\noindent $\bullet$
1500{\AA} attenuation and UV spectral slope $\beta$ do
correlate with each other (see Fig.\ \ref{fig:beta}), although with a large
scatter and a substantially steeper slope than was expected on the 
basis of the LMC extinction curve or the empirically-derived 
correlation between $A_{1500}$ and $\beta$ for starbursting 
galaxies \citep{calzetti94}.  Ultra-luminous infrared
galaxies populate yet another part of $A_{1500}$/$\beta$ space 
\citep{meurer00}.  While some scatter may be due to 
variations in foreground galactic extinction, 
both the large scatter and variation in slope
between a variety of star-forming systems demonstrates 
empirically that $\beta$ is not necessarily an accurate indicator 
of \fuv attenuation.  This point has been argued extensively 
from a theoretical perspective 
\citep[Fig.\ \protect\ref{fig:radtrans};][]{clump2}.
This conclusion carries with it serious implications for the
many studies of 
the rest frame UV properties of high-redshift galaxies that
utilize the local correlation between $\beta$ and \afif
for starbursting galaxies to (substantially) correct the observed 
UV fluxes for extinction. If, as is suggested by our work
and the work of \citet{meurer00}, the correlation between 
$\beta$ and \afif can have a variable slope and large scatter,
serious thought must be devoted to the suitability of 
using $\beta$ to correct UV fluxes for extinction.

\acknowledgements

We thank R.\ Cornett, J.\ Parker, M.\ Braun and 
M. Filipovi\'c for providing images of the LMC in the UV,
$R$-band, \firns, and radio respectively.  We also wish to 
thank the referee, Geoff Clayton, Daniella Calzetti and Claus Leitherer for
their useful comments.
E.\ F.\ B.\ and R.\ C.\ K.\ were supported
by NASA grant NAG5-8426 and NSF grant AST-9900789.  
K.\ D.\ G.\ was supported by NASA grants NAGW-3168 and NAG5-7933.
D.\ Z.\ was
supported by NASA grant NAG5-3501 and NSF grant AST-9619576.

\onecolumn
\clearpage

\begin{table}
\begin{scriptsize}
\begin{center}
\caption{\hii region fluxes through 4.9{\arcmin} apertures {\label{tab:flux}}} 
\begin{tabular}{lccccccccc}
\tableline
\tableline
Name & 1500{\AA} & 1900{\AA} & H$\alpha$ & 8.3$\micron$ &
12$\micron$ & 25$\micron$ & 60$\micron$ & 100$\micron$ &
8.55 GHz \\
 & (Jy) & (Jy) & (10$^{-10}$ ergs\,s$^{-1}$\,cm$^{-2}$) & (Jy)  &
(Jy) & (Jy) & (Jy) & (Jy) &
(Jy) \\
\tableline
 	N77E &  0.077 $\pm$  0.012 &  0.145 $\pm$  0.016 & 
   1.32 $\pm$    0.07 &    2.71 $\pm$    0.28 & 
   3.07 $\pm$    0.24 &   10.5 $\pm$    0.3 &   91.5 $\pm$    3.5 & 
 128.0 $\pm$   10.8 &   0.261 $\pm$   0.015 \\ 
 	N4AB &  0.057 $\pm$  0.014 &  0.075 $\pm$  0.017 & 
   0.99 $\pm$    0.05 &    4.36 $\pm$    0.11 & 
   5.13 $\pm$    0.06 &   16.0 $\pm$    0.1 &  102.9 $\pm$    0.8 & 
 151.5 $\pm$    2.5 &   0.430 $\pm$   0.022 \\ 
 	N79AB &  0.186 $\pm$  0.012 &  0.499 $\pm$  0.033 & 
   1.86 $\pm$    0.11 &    6.99 $\pm$    0.28 & 
   7.03 $\pm$    0.37 &   28.6 $\pm$    0.5 &  191.9 $\pm$    7.5 & 
 288.0 $\pm$   20.4 &   0.663 $\pm$   0.038 \\ 
 	N79DE &  0.317 $\pm$  0.022 &  0.559 $\pm$  0.045 & 
   2.70 $\pm$    0.14 &    4.42 $\pm$    0.19 & 
   5.04 $\pm$    0.38 &   18.0 $\pm$    0.5 &  164.9 $\pm$    7.1 & 
 261.4 $\pm$   17.2 &   0.544 $\pm$   0.030 \\ 
 	N81 &  0.067 $\pm$  0.015 &  0.146 $\pm$  0.033 & 
   0.61 $\pm$    0.12 &    2.86 $\pm$    0.16 & 
   3.37 $\pm$    0.25 &    6.9 $\pm$    0.4 &   61.2 $\pm$    7.2 & 
  98.6 $\pm$   12.6 &   0.115 $\pm$   0.022 \\ 
 	N83 &  0.297 $\pm$  0.019 &  0.526 $\pm$  0.034 & 
   2.77 $\pm$    0.14 &    6.59 $\pm$    0.19 & 
   7.66 $\pm$    0.18 &   28.3 $\pm$    0.3 &  232.5 $\pm$    4.9 & 
 381.7 $\pm$   11.4 &   0.611 $\pm$   0.033 \\ 
 	N11F &  1.280 $\pm$  0.065 &  1.599 $\pm$  0.082 & 
   1.85 $\pm$    0.12 &    2.84 $\pm$    0.33 & 
   3.46 $\pm$    0.48 &   10.8 $\pm$    0.7 &   99.3 $\pm$    9.9 & 
 142.9 $\pm$   23.8 &   0.338 $\pm$   0.029 \\ 
 	N11B &  1.073 $\pm$  0.054 &  1.319 $\pm$  0.067 & 
   7.95 $\pm$    0.40 &    7.73 $\pm$    0.19 & 
  10.52 $\pm$    0.20 &   61.3 $\pm$    0.3 &  392.7 $\pm$    2.8 & 
 486.5 $\pm$    5.3 &   1.994 $\pm$   0.102 \\ 
 	N91 &  0.706 $\pm$  0.035 &  0.997 $\pm$  0.050 & 
   2.86 $\pm$    0.14 &    3.66 $\pm$    0.09 & 
   4.47 $\pm$    0.09 &   19.4 $\pm$    0.2 &  164.1 $\pm$    2.5 & 
 219.0 $\pm$    6.7 &   0.521 $\pm$   0.031 \\ 
 	N11CD &  0.424 $\pm$  0.023 &  0.626 $\pm$  0.033 & 
   4.34 $\pm$    0.23 &    5.13 $\pm$    0.62 & 
   6.33 $\pm$    0.74 &   25.9 $\pm$    1.0 &  215.2 $\pm$   10.2 & 
 285.8 $\pm$   29.4 &   0.909 $\pm$   0.047 \\ 
 	N11E &  0.029 $\pm$  0.008 &  0.090 $\pm$  0.007 & 
   1.10 $\pm$    0.09 &    2.72 $\pm$    0.33 & 
   3.35 $\pm$    0.36 &    9.1 $\pm$    0.4 &   90.1 $\pm$    3.4 & 
 160.6 $\pm$   11.2 &   0.245 $\pm$   0.021 \\ 
 	N23A &  0.635 $\pm$  0.033 &  0.984 $\pm$  0.050 & 
   0.95 $\pm$    0.05 &    1.17 $\pm$    0.11 & 
   1.26 $\pm$    0.07 &    3.7 $\pm$    0.1 &   40.6 $\pm$    0.7 & 
  76.3 $\pm$    1.8 &   0.211 $\pm$   0.011 \\ 
 	N103B &  0.787 $\pm$  0.039 &  0.903 $\pm$  0.046 & 
   1.19 $\pm$    0.06 &    1.62 $\pm$    0.24 & 
   1.52 $\pm$    0.30 &    3.2 $\pm$    0.4 &   45.5 $\pm$    5.6 & 
  73.1 $\pm$   13.7 &   0.246 $\pm$   0.016 \\ 
 	N105A &  0.530 $\pm$  0.027 &  0.688 $\pm$  0.035 & 
   3.92 $\pm$    0.20 &   10.67 $\pm$    0.45 & 
  12.56 $\pm$    0.36 &   45.3 $\pm$    0.4 &  359.3 $\pm$    4.6 & 
 524.1 $\pm$   13.8 &   0.966 $\pm$   0.050 \\ 
 	N113 S &  0.764 $\pm$  0.041 &  0.914 $\pm$  0.053 & 
   2.75 $\pm$    0.14 &    9.10 $\pm$    0.36 & 
  10.60 $\pm$    0.31 &   41.7 $\pm$    0.4 &  304.5 $\pm$    7.4 & 
 458.9 $\pm$   20.3 &   0.629 $\pm$   0.037 \\ 
 	N113 N &  0.586 $\pm$  0.030 &  0.809 $\pm$  0.042 & 
   1.71 $\pm$    0.09 &    2.25 $\pm$    0.19 & 
   2.49 $\pm$    0.14 &    6.2 $\pm$    0.3 &   86.1 $\pm$    5.0 & 
 148.5 $\pm$    6.6 &   0.251 $\pm$   0.023 \\ 
 	N30BC &  0.181 $\pm$  0.009 &  0.335 $\pm$  0.017 & 
   1.40 $\pm$    0.07 &    3.17 $\pm$    0.08 & 
   3.07 $\pm$    0.05 &    8.7 $\pm$    0.1 &   82.8 $\pm$    0.9 & 
 130.1 $\pm$    3.0 &   0.132 $\pm$   0.015 \\ 
 	N119 &  2.129 $\pm$  0.109 &  1.707 $\pm$  0.096 & 
   3.37 $\pm$    0.17 &    4.59 $\pm$    0.34 & 
   5.14 $\pm$    0.42 &   22.0 $\pm$    0.7 &  192.9 $\pm$   13.2 & 
 237.3 $\pm$   26.8 &   0.503 $\pm$   0.028 \\ 
 	N120ABC &  0.720 $\pm$  0.050 &  0.817 $\pm$  0.079 & 
   2.34 $\pm$    0.12 &    8.00 $\pm$    0.28 & 
   8.92 $\pm$    0.31 &   26.6 $\pm$    0.2 &  254.4 $\pm$    4.1 & 
 327.7 $\pm$   13.5 &   0.523 $\pm$   0.030 \\ 
 	N44BC &  1.408 $\pm$  0.070 &  1.262 $\pm$  0.063 & 
   4.15 $\pm$    0.21 &    8.85 $\pm$    0.19 & 
  11.16 $\pm$    0.24 &   54.0 $\pm$    0.2 &  375.7 $\pm$    2.9 & 
 516.3 $\pm$    8.9 &   0.970 $\pm$   0.051 \\ 
 	N44I &  0.490 $\pm$  0.025 &  0.686 $\pm$  0.035 & 
   1.70 $\pm$    0.09 &    3.86 $\pm$    0.28 & 
   4.75 $\pm$    0.28 &   15.7 $\pm$    0.4 &  172.0 $\pm$    4.9 & 
 227.0 $\pm$   12.0 &   0.346 $\pm$   0.039 \\ 
 	N44D &  0.140 $\pm$  0.007 &  0.319 $\pm$  0.016 & 
   1.90 $\pm$    0.10 &    4.96 $\pm$    0.25 & 
   6.03 $\pm$    0.31 &   19.9 $\pm$    0.3 &  188.4 $\pm$    6.1 & 
 261.0 $\pm$   15.9 &   0.433 $\pm$   0.026 \\ 
 	N138A &  0.098 $\pm$  0.007 &  0.156 $\pm$  0.011 & 
   0.83 $\pm$    0.04 &    2.97 $\pm$    0.21 & 
   3.15 $\pm$    0.17 &    6.5 $\pm$    0.3 &   68.0 $\pm$    6.3 & 
 130.5 $\pm$   14.6 &   0.108 $\pm$   0.015 \\ 
 	N48AC &  0.209 $\pm$  0.011 &  0.318 $\pm$  0.019 & 
   0.93 $\pm$    0.05 &    5.02 $\pm$    0.15 & 
   5.80 $\pm$    0.16 &   16.7 $\pm$    0.3 &  146.5 $\pm$    4.5 & 
 214.0 $\pm$   12.3 &   0.343 $\pm$   0.026 \\ 
 	N51D &  1.332 $\pm$  0.067 &  1.052 $\pm$  0.054 & 
   2.24 $\pm$    0.11 &    0.47 $\pm$    0.11 & 
   0.80 $\pm$    0.08 &    3.7 $\pm$    0.1 &   44.3 $\pm$    2.4 & 
  59.8 $\pm$    3.6 &   0.316 $\pm$   0.026 \\ 
 	N51E &  0.153 $\pm$  0.011 &  0.254 $\pm$  0.015 & 
   1.02 $\pm$    0.07 &    0.70 $\pm$    0.14 & 
   0.90 $\pm$    0.16 &    2.7 $\pm$    0.2 &   38.3 $\pm$    3.9 & 
  57.1 $\pm$    7.7 &   0.186 $\pm$   0.017 \\ 
 	N143 &  0.256 $\pm$  0.027 &  0.295 $\pm$  0.043 & 
   0.72 $\pm$    0.06 &    0.43 $\pm$    0.17 & 
   0.77 $\pm$    0.29 &    2.6 $\pm$    0.5 &   30.7 $\pm$    4.2 & 
  58.2 $\pm$    9.8 &   0.140 $\pm$   0.031 \\ 
 	N144AB &  2.000 $\pm$  0.101 &  1.343 $\pm$  0.074 & 
   3.72 $\pm$    0.20 &    5.62 $\pm$    0.16 & 
   7.28 $\pm$    0.15 &   26.6 $\pm$    0.2 &  216.6 $\pm$    4.0 & 
 263.7 $\pm$    7.7 &   0.538 $\pm$   0.030 \\ 
 	N51C &  1.145 $\pm$  0.058 &  1.075 $\pm$  0.060 & 
   1.98 $\pm$    0.10 &    2.25 $\pm$    0.11 & 
   2.85 $\pm$    0.10 &   13.0 $\pm$    0.2 &  122.0 $\pm$    3.2 & 
 171.8 $\pm$    6.5 &   0.263 $\pm$   0.027 \\ 
 	N51A &  1.048 $\pm$  0.058 &  0.941 $\pm$  0.058 & 
   1.46 $\pm$    0.08 &    2.77 $\pm$    0.10 & 
   2.94 $\pm$    0.08 &    8.8 $\pm$    0.1 &  104.1 $\pm$    2.5 & 
 165.2 $\pm$    4.8 &   0.222 $\pm$   0.021 \\ 
 	N148 &  0.250 $\pm$  0.018 &  0.342 $\pm$  0.026 & 
   0.94 $\pm$    0.08 &    2.54 $\pm$    0.21 & 
   2.79 $\pm$    0.20 &    5.1 $\pm$    0.2 &   76.4 $\pm$    3.8 & 
 133.5 $\pm$    8.9 &   0.172 $\pm$   0.012 \\ 
 	N55 N &  0.460 $\pm$  0.024 &  0.578 $\pm$  0.033 & 
   1.38 $\pm$    0.07 &    2.78 $\pm$    0.06 & 
   2.95 $\pm$    0.02 &    8.2 $\pm$    0.1 &   96.0 $\pm$    0.2 & 
 164.9 $\pm$    0.7 &   0.219 $\pm$   0.014 \\ 
 	N55A &  0.718 $\pm$  0.036 &  0.760 $\pm$  0.039 & 
   2.00 $\pm$    0.10 &    2.84 $\pm$    0.09 & 
   3.29 $\pm$    0.03 &   12.5 $\pm$    0.0 &  118.2 $\pm$    0.2 & 
 175.4 $\pm$    0.4 &   0.228 $\pm$   0.020 \\ 
 	N57A &  1.630 $\pm$  0.085 &  1.407 $\pm$  0.080 & 
   2.85 $\pm$    0.14 &    6.19 $\pm$    0.11 & 
   7.29 $\pm$    0.14 &   23.5 $\pm$    0.2 &  232.7 $\pm$    4.9 & 
 334.7 $\pm$   10.8 &   0.434 $\pm$   0.024 \\ 
 	Fil A &  0.252 $\pm$  0.039 &  0.410 $\pm$  0.078 & 
   0.68 $\pm$    0.07 &    0.95 $\pm$    0.16 & 
   1.24 $\pm$    0.13 &    2.5 $\pm$    0.2 &   40.5 $\pm$    2.7 & 
  71.8 $\pm$    5.3 &   0.052 $\pm$   0.024 \\ 
 	N63A &  0.455 $\pm$  0.023 &  0.551 $\pm$  0.028 & 
   1.02 $\pm$    0.06 &    1.57 $\pm$    0.13 & 
   1.54 $\pm$    0.07 &    6.4 $\pm$    0.1 &   58.8 $\pm$    1.2 & 
  91.2 $\pm$    3.4 &   0.452 $\pm$   0.024 \\ 
 	N59A &  0.283 $\pm$  0.017 &  0.359 $\pm$  0.025 & 
   5.44 $\pm$    0.27 &    7.66 $\pm$    0.14 & 
   9.34 $\pm$    0.14 &   45.9 $\pm$    0.2 &  327.7 $\pm$    2.8 & 
 443.1 $\pm$    5.7 &   1.376 $\pm$   0.071 \\ 
 	N154A &  0.344 $\pm$  0.023 &  0.454 $\pm$  0.037 & 
   2.99 $\pm$    0.24 &    6.35 $\pm$    0.42 & 
   7.27 $\pm$    0.41 &   23.3 $\pm$    0.3 &  257.9 $\pm$   14.6 & 
 365.9 $\pm$   29.8 &   0.521 $\pm$   0.045 \\ 
 	N64AB &  0.205 $\pm$  0.019 &  0.248 $\pm$  0.023 & 
   1.07 $\pm$    0.06 &    2.25 $\pm$    0.12 & 
   2.16 $\pm$    0.11 &    4.6 $\pm$    0.1 &   55.9 $\pm$    2.2 & 
 108.6 $\pm$    6.2 &   0.194 $\pm$   0.014 \\ 
 	N158C &  0.913 $\pm$  0.062 &  0.840 $\pm$  0.095 & 
   6.04 $\pm$    0.43 &   10.56 $\pm$    0.57 & 
  13.40 $\pm$    0.77 &   72.0 $\pm$    1.6 &  538.8 $\pm$   33.6 & 
 665.2 $\pm$   54.8 &   1.138 $\pm$   0.064 \\ 
 	N160AD &  0.262 $\pm$  0.027 &  0.435 $\pm$  0.055 & 
   7.36 $\pm$    0.39 &   21.07 $\pm$    0.72 & 
  26.03 $\pm$    0.94 &  120.3 $\pm$    2.3 &  815.5 $\pm$   34.7 & 
 970.3 $\pm$   62.8 &   2.418 $\pm$   0.123 \\ 
 	N158 &  0.341 $\pm$  0.019 &  0.524 $\pm$  0.036 & 
   2.43 $\pm$    0.47 &    2.44 $\pm$    0.65 & 
   3.18 $\pm$    0.72 &   15.3 $\pm$    3.5 &  140.2 $\pm$   45.1 & 
 151.2 $\pm$   61.8 &   0.435 $\pm$   0.104 \\ 
 	N159 &  0.099 $\pm$  0.023 &  0.271 $\pm$  0.041 & 
   6.21 $\pm$    0.32 &   21.44 $\pm$    0.72 & 
  30.73 $\pm$    0.95 &  170.0 $\pm$    1.7 & 1153.3 $\pm$   24.8 & 
1253.1 $\pm$   53.1 &   2.808 $\pm$   0.142 \\ 
 	N158A &  0.363 $\pm$  0.026 &  0.493 $\pm$  0.040 & 
   1.33 $\pm$    0.65 &    0.16 $\pm$    0.83 & 
   0.61 $\pm$    1.11 &    1.8 $\pm$    4.0 &   10.6 $\pm$   46.7 & 
   1.7 $\pm$   72.5 &   0.218 $\pm$   0.137 \\ 
 	N160BCE &  0.307 $\pm$  0.021 &  0.439 $\pm$  0.034 & 
   4.51 $\pm$    0.24 &    6.79 $\pm$    0.64 & 
   8.14 $\pm$    0.87 &   36.9 $\pm$    1.8 &  369.6 $\pm$   26.9 & 
 526.6 $\pm$   60.4 &   1.026 $\pm$   0.051 \\ 
 	N175 &  0.020 $\pm$  0.003 &  0.023 $\pm$  0.003 & 
   0.47 $\pm$    0.03 &    0.26 $\pm$    0.75 & 
   0.52 $\pm$    0.98 &    1.1 $\pm$    0.9 &   19.2 $\pm$   15.6 & 
  32.8 $\pm$   46.1 &   0.123 $\pm$   0.017 \\ 
 	N214C &  0.077 $\pm$  0.005 &  0.145 $\pm$  0.011 & 
   2.28 $\pm$    0.11 &    1.13 $\pm$    0.17 & 
   1.98 $\pm$    0.12 &    8.2 $\pm$    0.1 &   80.2 $\pm$    0.9 & 
 130.4 $\pm$    3.7 &   0.348 $\pm$   0.031 \\ 
 	NGC 2100 &  0.338 $\pm$  0.024 &  0.583 $\pm$  0.042 & 
   0.43 $\pm$    0.23 &    1.31 $\pm$    0.46 & 
   1.46 $\pm$    0.47 &    2.0 $\pm$    1.5 &   25.7 $\pm$   23.2 & 
  46.0 $\pm$   46.7 &   0.093 $\pm$   0.042 \\ 
 	N164 &  0.093 $\pm$  0.019 &  0.182 $\pm$  0.037 & 
   2.34 $\pm$    0.21 &    2.62 $\pm$    0.58 & 
   3.25 $\pm$    0.54 &   14.3 $\pm$    1.1 &  163.4 $\pm$   15.3 & 
 210.7 $\pm$   35.7 &   0.492 $\pm$   0.061 \\ 
 	N165 &  0.043 $\pm$  0.011 &  0.083 $\pm$  0.017 & 
   0.54 $\pm$    0.08 &    1.05 $\pm$    0.26 & 
   1.12 $\pm$    0.22 &    2.8 $\pm$    0.4 &   64.8 $\pm$    3.1 & 
  95.5 $\pm$   10.5 &   0.103 $\pm$   0.037 \\ 
 	N163 &  0.019 $\pm$  0.007 &  0.056 $\pm$  0.010 & 
   1.27 $\pm$    0.14 &    3.16 $\pm$    0.29 & 
   3.82 $\pm$    0.30 &    8.6 $\pm$    0.6 &   93.5 $\pm$   10.4 & 
 160.9 $\pm$   23.7 &   0.227 $\pm$   0.046 \\ 
 	N180AB &  0.357 $\pm$  0.018 &  0.515 $\pm$  0.026 & 
   3.44 $\pm$    0.17 &    2.90 $\pm$    0.11 & 
   3.19 $\pm$    0.22 &   13.3 $\pm$    0.2 &  120.6 $\pm$    2.6 & 
 199.2 $\pm$   11.3 &   0.577 $\pm$   0.032 \\ 
\tableline \\
\end{tabular}
\end{center}
\end{scriptsize}
\end{table}

\clearpage

\begin{table}
\begin{scriptsize}
\begin{center}
\caption{\hii region attenuation estimates and UV spectral slopes through 4.9{\arcmin} apertures {\label{tab:atten}}} 
\begin{tabular}{lccccc}
\tableline
\tableline
Name & $A_{{\rm H\alpha,Balmer}}$ & $A_{{\rm H\alpha,Radio}}$ & 
$A_{{\rm 1500,FR}}$ & $A_{{\rm 1900,FR}}$ & $\beta$   \\
 & (mag) & (mag) & (mag) & (mag) & \\
\tableline
 	N77E &  0.31 $\pm$  0.09 &  0.62 $\pm$  0.11 & 
 1.17 $\pm$  0.21 &  0.72 $\pm$  0.20 &  0.48 $\pm$  0.72 \\ 
 	N4AB &  0.12 $\pm$  0.09 &  1.48 $\pm$  0.11 & 
 1.55 $\pm$  0.32 &  1.28 $\pm$  0.30 & -0.92 $\pm$  1.28 \\ 
 	N79AB &  0.47 $\pm$  0.09 &  1.26 $\pm$  0.11 & 
 1.13 $\pm$  0.17 &  0.52 $\pm$  0.16 &  1.85 $\pm$  0.36 \\ 
 	N79DE &  0.29 $\pm$  0.09 &  0.64 $\pm$  0.11 & 
 0.69 $\pm$  0.16 &  0.39 $\pm$  0.13 &  0.24 $\pm$  0.41 \\ 
 	N81 &  0.39 $\pm$  0.09 &  0.57 $\pm$  0.30 & 
 1.08 $\pm$  0.25 &  0.57 $\pm$  0.22 &  1.07 $\pm$  1.25 \\ 
 	N83 &  0.46 $\pm$  0.09 &  0.74 $\pm$  0.11 & 
 0.96 $\pm$  0.17 &  0.58 $\pm$  0.17 &  0.25 $\pm$  0.35 \\ 
 	N11F &  0.07 $\pm$  0.09 &  0.54 $\pm$  0.14 & 
 0.09 $\pm$  0.05 &  0.07 $\pm$  0.05 & -1.11 $\pm$  0.28 \\ 
 	N11B &  0.20 $\pm$  0.09 &  0.88 $\pm$  0.10 & 
 0.49 $\pm$  0.13 &  0.37 $\pm$  0.12 & -1.17 $\pm$  0.28 \\ 
 	N91 &  0.24 $\pm$  0.09 &  0.53 $\pm$  0.11 & 
 0.31 $\pm$  0.09 &  0.21 $\pm$  0.09 & -0.63 $\pm$  0.28 \\ 
 	N11CD &  0.10 $\pm$  0.09 &  0.68 $\pm$  0.11 & 
 0.64 $\pm$  0.15 &  0.42 $\pm$  0.13 & -0.46 $\pm$  0.30 \\ 
 	N11E &  0.20 $\pm$  0.09 &  0.75 $\pm$  0.15 & 
 2.01 $\pm$  0.41 &  1.06 $\pm$  0.20 &  2.39 $\pm$  1.15 \\ 
 	N23A &  0.15 $\pm$  0.09 &  0.75 $\pm$  0.11 & 
 0.08 $\pm$  0.05 &  0.05 $\pm$  0.05 & -0.27 $\pm$  0.28 \\ 
 	N103B &  0.12 $\pm$  0.09 &  0.67 $\pm$  0.11 & 
 0.06 $\pm$  0.05 &  0.06 $\pm$  0.05 & -1.44 $\pm$  0.28 \\ 
 	N105A &  0.15 $\pm$  0.09 &  0.86 $\pm$  0.11 & 
 0.84 $\pm$  0.17 &  0.64 $\pm$  0.17 & -0.95 $\pm$  0.28 \\ 
 	N113 S &  0.06 $\pm$  0.09 &  0.78 $\pm$  0.11 & 
 0.56 $\pm$  0.14 &  0.45 $\pm$  0.14 & -1.27 $\pm$  0.31 \\ 
 	N113 N &  0.14 $\pm$  0.09 &  0.29 $\pm$  0.13 & 
 0.20 $\pm$  0.07 &  0.14 $\pm$  0.07 & -0.72 $\pm$  0.28 \\ 
 	N30BC & \nodata & -0.18 $\pm$  0.15 & 
 0.63 $\pm$  0.15 &  0.34 $\pm$  0.12 &  0.42 $\pm$  0.28 \\ 
 	N119 & -0.03 $\pm$  0.09 &  0.32 $\pm$  0.11 & 
 0.10 $\pm$  0.05 &  0.13 $\pm$  0.07 & -2.83 $\pm$  0.29 \\ 
 	N120ABC &  0.05 $\pm$  0.09 &  0.76 $\pm$  0.11 & 
 0.46 $\pm$  0.12 &  0.38 $\pm$  0.13 & -1.48 $\pm$  0.46 \\ 
 	N44BC &  0.18 $\pm$  0.09 &  0.80 $\pm$  0.11 & 
 0.37 $\pm$  0.11 &  0.39 $\pm$  0.13 & -2.40 $\pm$  0.28 \\ 
 	N44I &  0.05 $\pm$  0.09 &  0.65 $\pm$  0.15 & 
 0.44 $\pm$  0.12 &  0.30 $\pm$  0.11 & -0.67 $\pm$  0.28 \\ 
 	N44D &  0.25 $\pm$  0.09 &  0.78 $\pm$  0.11 & 
 1.23 $\pm$  0.17 &  0.67 $\pm$  0.18 &  1.21 $\pm$  0.29 \\ 
 	N138A &  0.25 $\pm$  0.09 &  0.17 $\pm$  0.17 & 
 0.94 $\pm$  0.18 &  0.62 $\pm$  0.17 & -0.19 $\pm$  0.40 \\ 
 	N48AC &  0.46 $\pm$  0.09 &  1.30 $\pm$  0.12 & 
 0.86 $\pm$  0.17 &  0.58 $\pm$  0.16 & -0.35 $\pm$  0.31 \\ 
 	N51D & -0.07 $\pm$  0.09 &  0.26 $\pm$  0.13 & 
 0.02 $\pm$  0.04 &  0.04 $\pm$  0.04 & -2.89 $\pm$  0.28 \\ 
 	N51E &  0.17 $\pm$  0.09 &  0.53 $\pm$  0.14 & 
 0.32 $\pm$  0.10 &  0.18 $\pm$  0.08 & -0.00 $\pm$  0.36 \\ 
 	N143 &  0.05 $\pm$  0.09 &  0.60 $\pm$  0.27 & 
 0.17 $\pm$  0.07 &  0.14 $\pm$  0.08 & -1.43 $\pm$  0.70 \\ 
 	N144AB &  0.27 $\pm$  0.09 &  0.28 $\pm$  0.11 & 
 0.13 $\pm$  0.06 &  0.20 $\pm$  0.08 & -3.52 $\pm$  0.29 \\ 
 	N51C &  0.02 $\pm$  0.09 &  0.19 $\pm$  0.14 & 
 0.13 $\pm$  0.06 &  0.14 $\pm$  0.07 & -2.22 $\pm$  0.29 \\ 
 	N51A &  0.10 $\pm$  0.09 &  0.34 $\pm$  0.14 & 
 0.13 $\pm$  0.06 &  0.14 $\pm$  0.07 & -2.40 $\pm$  0.32 \\ 
 	N148 &  0.19 $\pm$  0.09 &  0.54 $\pm$  0.13 & 
 0.44 $\pm$  0.12 &  0.30 $\pm$  0.11 & -0.75 $\pm$  0.41 \\ 
 	N55 N &  0.11 $\pm$  0.09 &  0.38 $\pm$  0.11 & 
 0.30 $\pm$  0.09 &  0.23 $\pm$  0.09 & -1.09 $\pm$  0.30 \\ 
 	N55A &  0.10 $\pm$  0.09 &  0.02 $\pm$  0.13 & 
 0.22 $\pm$  0.08 &  0.20 $\pm$  0.08 & -1.75 $\pm$  0.28 \\ 
 	N57A &  0.23 $\pm$  0.09 &  0.34 $\pm$  0.11 & 
 0.19 $\pm$  0.07 &  0.21 $\pm$  0.09 & -2.54 $\pm$  0.30 \\ 
 	Fil A & -0.06 $\pm$  0.09 & -0.41 $\pm$  0.51 & 
 0.23 $\pm$  0.09 &  0.13 $\pm$  0.08 & -0.08 $\pm$  0.95 \\ 
 	N63A &  0.08 $\pm$  0.09 &  1.49 $\pm$  0.11 & 
 0.18 $\pm$  0.07 &  0.14 $\pm$  0.07 & -1.23 $\pm$  0.28 \\ 
 	N59A &  0.50 $\pm$  0.09 &  0.89 $\pm$  0.11 & 
 1.15 $\pm$  0.17 &  0.94 $\pm$  0.20 & -1.05 $\pm$  0.36 \\ 
 	N154A &  0.19 $\pm$  0.09 &  0.49 $\pm$  0.15 & 
 0.85 $\pm$  0.17 &  0.64 $\pm$  0.18 & -0.89 $\pm$  0.41 \\ 
 	N64AB &  0.11 $\pm$  0.09 &  0.53 $\pm$  0.12 & 
 0.43 $\pm$  0.13 &  0.33 $\pm$  0.12 & -1.24 $\pm$  0.51 \\ 
 	N158C &  0.37 $\pm$  0.09 &  0.57 $\pm$  0.12 & 
 0.71 $\pm$  0.16 &  0.70 $\pm$  0.20 & -2.30 $\pm$  0.51 \\ 
 	N160AD &  0.44 $\pm$  0.09 &  1.17 $\pm$  0.11 & 
 1.90 $\pm$  0.26 &  1.41 $\pm$  0.24 & -0.00 $\pm$  0.63 \\ 
 	N158 &  0.19 $\pm$  0.09 &  0.51 $\pm$  0.34 & 
 0.48 $\pm$  0.13 &  0.30 $\pm$  0.11 & -0.30 $\pm$  0.35 \\ 
 	N159 &  0.65 $\pm$  0.09 &  1.52 $\pm$  0.11 & 
 3.11 $\pm$  0.42 &  2.10 $\pm$  0.30 &  1.92 $\pm$  1.06 \\ 
 	N158A &  0.16 $\pm$  0.09 &  0.42 $\pm$  0.87 & 
 0.01 $\pm$  0.04 &  0.01 $\pm$  0.04 & -0.78 $\pm$  0.42 \\ 
 	N160BCE &  0.27 $\pm$  0.09 &  0.77 $\pm$  0.11 & 
 1.14 $\pm$  0.17 &  0.86 $\pm$  0.20 & -0.59 $\pm$  0.40 \\ 
 	N175 &  0.31 $\pm$  0.09 &  0.94 $\pm$  0.18 & 
 1.04 $\pm$  0.20 &  0.87 $\pm$  0.22 & -1.31 $\pm$  0.69 \\ 
 	N214C &  0.15 $\pm$  0.09 &  0.34 $\pm$  0.13 & 
 1.10 $\pm$  0.17 &  0.66 $\pm$  0.18 &  0.50 $\pm$  0.39 \\ 
 	NGC 2100 &  0.08 $\pm$  0.09 &  0.73 $\pm$  0.76 & 
 0.11 $\pm$  0.06 &  0.06 $\pm$  0.05 &  0.14 $\pm$  0.39 \\ 
 	N164 &  0.50 $\pm$  0.09 &  0.69 $\pm$  0.18 & 
 1.34 $\pm$  0.26 &  0.86 $\pm$  0.25 &  0.61 $\pm$  1.11 \\ 
 	N165 &  0.25 $\pm$  0.09 &  0.58 $\pm$  0.42 & 
 1.24 $\pm$  0.29 &  0.77 $\pm$  0.24 &  0.60 $\pm$  1.32 \\ 
 	N163 &  0.53 $\pm$  0.09 &  0.51 $\pm$  0.26 & 
 2.43 $\pm$  0.52 &  1.40 $\pm$  0.28 &  2.24 $\pm$  1.62 \\ 
 	N180AB &  0.16 $\pm$  0.09 &  0.44 $\pm$  0.11 & 
 0.48 $\pm$  0.12 &  0.32 $\pm$  0.11 & -0.55 $\pm$  0.28 \\ 
\tableline \\
\end{tabular}
\end{center}
\end{scriptsize}
\end{table}

\end{document}